\documentstyle[12pt]{article}
\begin{document}
\baselineskip=0.7cm
\newcommand{\EQ}{\begin{equation}}
\newcommand{\EN}{\end{equation}}
\newcommand{\EQA}{\begin{eqnarray}}
\newcommand{\EQN}{\end{eqnarray}}
\newcommand{\e}{{\rm e}}
\newcommand{\Sp}{{\rm Sp}}
\renewcommand{\theequation}{\arabic{section}.\arabic{equation}}
\newcommand{\Tr}{{\rm Tr}}
\renewcommand{\thesection}{\arabic{section}.}
\renewcommand{\thesubsection}{\arabic{section}.\arabic{subsection}}
\makeatletter
\def\section{\@startsection{section}{1}{\z@}{-3.5ex plus -1ex minus 
 -.2ex}{2.3ex plus .2ex}{\large}} 
\def\subsection{\@startsection{subsection}{2}{\z@}{-3.25ex plus -1ex minus 
 -.2ex}{1.5ex plus .2ex}{\normalsize\it}}
\def\appendix{
\par
\setcounter{section}{0}
\setcounter{subsection}{0}
\def\thesection{\Alph{section}}}
\makeatother
\def\thefootnote{\fnsymbol{footnote}}
\begin{flushright}
UT-KOMABA/98-13\\
June 1998
\end{flushright}
\vspace{1cm}
\begin{center}
\Large
Multi-Body Interactions of D-Particles \\
in Supergravity and 
Matrix Theory

\vspace{1cm}
\normalsize
{\sc Yuji Okawa}
\footnote{
e-mail address:\ \ okawa@hep1.c.u-tokyo.ac.jp}
and 
{\sc Tamiaki Yoneya}
\footnote{
e-mail address:\ \ tam@hep1.c.u-tokyo.ac.jp}
\\
\vspace{0.3cm}
 {\it Institute of Physics, University of Tokyo, Komaba, Tokyo}

\vspace{1.3cm}
Abstract\\

\end{center}
We present  detailed analyses of the 3-body interactions 
of D-particles from both sides of 11 dimensional 
supergravity and Matrix theory. 
In supergravity, we derive a complete  
expression for the classical bosonic effective action for D-particles 
including 2-and 3-body 
interaction terms. In Matrix theory, we compute 
 1-particle irreducible 
contributions to the 
eikonal phase shift in the two-loop approximation. 
The results precisely agree with the predictions from 
supergravity and thus provide a strong support 
to the discrete light-cone interpretation of the 
Matrix-theory conjecture as 
a possible nonperturbative definition of M-theory. 

\newpage
\section{Introduction}
It has recently been hoped that D-branes \cite{pol} provide a 
keystone towards long-sought nonperturbative reformulation of 
string theory.  In particular,  the point-like 
D-branes, namely D-particles,  may be 
the fundamental degrees of freedom from the viewpoint 
of the so-called M-theory \cite{witten} which has been proposed as 
the non-perturbative correspondent of type IIA string theory. 
In the low-energy approximation,  M-theory 
is expected to be described by 11 dimensional supergravity. 
Since the initial proposal of Matrix theory \cite{bfss} 
by Banks, Fischler, Shenker and Susskind as a 
possible realization of the above general conjecture, 
an impressive amount of evidence concerning its relation with 
eleven-dimensional supergravity has been accumulated.  

However,  there has also been reported several results 
(see, e. g. \cite{douglas}\cite{ganor}\cite{dine}\cite{keski}) 
which suggest 
discrepancies between Matrix theory and supergravity.  
Among such claims, one of the most serious problems seems 
to be that of the multi-body interactions \cite{dine} of D-particles. 
In view of the agreement \cite{bb}\cite{bbpt} of the finite $N$ two-loop contribution 
for the two-body interaction of D-particles 
with that of supergravity, 
we should naturally expect that  supergravity  
should be consistent with Matrix theory 
for 3-body interactions too, 
since the latter are of the same order of magnitude as the 
two-body interactions, 
$N^3 {v^6 \over r^{14}}$,  
with respect to all the external variables involved, 
the masses or the $N$, the relative velocities and 
distances between the D-particles.   
In fact, however, there are important subtleties. 
The emergence of correct results is not 
automatic in Matrix theory which is 
only formulated in the light-cone or infinite-momentum 
frame.  If D-particle is interpreted from the viewpoint of 
 10 dimensional
 type IIA string theory, the leading two-body interaction 
cannot detect the non-linear nature of gravity.  
Because of the BPS property, it turns out that 
the self-interaction of gravitons does not contribute 
to the two-body force 
even in the two-loop (and higher) approximation.  
Therefore, the leading 
3-body interaction is the first instance where 
the self-interaction of gravitons enters. 
In this sense, the 3-body interaction can be regarded as a  
crucial testing ground where we can check 
whether the theory becomes the correct non-linear theory 
as required by Lorentz invariance, 
since, as is well known, we can reasonably expect that 
the consistent Lorentz invariant theories of interacting 
massless spin two particles necessarily include the Einstein gravity 
at least in the low-energy regime. 

From the viewpoint of 11D supergravity, on the 
other hand, the leading 2-body interaction itself involves the 
self-interaction of gravitons, since a 
D-particle is assumed 
to be nothing but a particular Kaluza-Klein mode of 
eleven-dimensional massless graviton.  
The Lorentz invariance would therefore require 
that the self-interaction of gravitons in non Kaluza-Klein modes 
should be correctly reproduced as well. 
In other words, the agreement of the 3-body interaction 
between supergravity and Matrix theory  
 would provide strong evidence for the Lorentz invariance 
of Matrix theory from both the 10 and 11 dimensional standpoints.  

Motivated by these considerations, we decided to carry out a 
detailed  investigation of the 3-body interactions of 
D-particles from the viewpoints of both  11D 
supergravity and the discrete light-cone formulation of 
Matrix theory. The present paper is the first report of 
our work in this direction.  Contrary to the previous claim 
in \cite{dine}, the exact agreement between 
 supergravity and Matrix theory is found. 

In section 2, we present an analysis of the 
interactions of D-particles in 11D supergravity 
in the classical approximation.  
The goal is to derive the effective 
action, based on which we can compute the 
scattering phase shift of D-particles in the 
eikonal approximation.  
In section 3, the computation of the scattering phase shift 
in the two-loop approximation in Matrix theory 
is carried out.  We present a complete closed formula for 
1-particle irreducible contributions to the 
phase shift. The result is an extension of the previous 
results reported in ref. \cite{bb}\cite{bbpt} for two-body interactions.  
In section 4, the comparison between  supergravity 
and  Matrix theory is made by explicitly computing 
the eikonal phase shift using the effective action obtained in 
section 2.   Finally, in section 5, we conclude 
with some remarks.  Some of detailed formulas 
are presented in Appendices. 
 
\section{D-Particle effective action in supergravity}

\subsection{Gravitational fields  of D-particles}

We first present a detailed derivation of the 
effective action for D-particle interactions from 
the viewpoint of eleven dimensional supergravity. 
A D-particle (more precisely the  transverse-scalar part of 
a graviton wave packet) which is 
at rest in the 9 dimensional transverse space is then interpreted as a 
``pencil of gravitational wave" in 11 dimensional space-time 
whose trajectory is given by $x^{+}=0 $ 
with $N$ units of the Kaluza-Klein momentum $p_-=N/R$ in the 
null 11th direction $x^{-}$ which is 
 compactified to a circle of radius $R$.  
We can effectively represent the D-particle 
as a singular exact solution of supergravity corresponding to a 
gravitational shock wave.   
Upon the reduction to 10 dimensions, 
it coincides with the exact (but singular at the 
origin) BPS solution of 
10 dimensional type IIA supergravity, with correct 
coupling with the dilaton and the 1-form R-R gauge 
field. Since our aim is the comparison with  two-loop calculations 
in Matrix theory  which corresponds to the 
tree contribution in  supergravity, it is sufficient to treat only the 
purely metric part of  supergravity.  
In this approximation, we can neglect the 
gravitinos and the 3-form field. 

Therefore,  we can start from the standard Einstein action 
in eleven dimensions 
\EQ
S = S_g + S_D ,
\label{eq21}
\EN
\EQ
\label{eq22}
S_g ={1\over 2\kappa_{11}^2}\int d^{11}x \sqrt{-g}R
\EN
by adding the auxiliary action $S_D$ which takes into account  singular (i.e., $\delta$-function) nature of the D-particle  solutions.  
The basic equation is therefore the Einstein equation in 
11 dimensional space-time.
\EQ
G_{\mu\lambda} \equiv R_{\mu\lambda} 
-{1\over 2}g_{\mu\lambda} R = \kappa_{11}^2 T_{\mu\lambda} . 
\label{eq23}
\EN
We use the same convention as ref \cite{bbpt} for the Newton constant $\kappa^2 = 16\pi^5 /M^9$ with $M$ being the 
11 dimensional Planck mass. 
The energy-momentum tensor for a D-particle 
can be assumed to be
\EQ
T^{\mu\lambda}(x) =
K \int ds \, \lambda(s)
{dx^{\mu}(s)\over ds}{dx^{\lambda}(s) \over ds}
{1\over \sqrt{-g(x(s))}}\delta (x-x(s)) 
\label{eq24}
\EN
corresponding to the action for a massless point particle 
\EQ
S_p = {K\over 2}\int \, ds \, \lambda(s)\,  
g_{\mu\lambda}(x(s) )\, 
{dx^{\mu}(s)\over ds}{dx^{\lambda}(s) \over ds}
\EN
where $\lambda(s)$ is the Lagrange multiplier 
imposing the massless constraint
\EQ
g_{\mu\lambda}(x(s) )\, 
{dx^{\mu}(s)\over ds}{dx^{\lambda}(s) \over ds}
=0
\label{eq26}
\EN
and $K$ is an arbitrary constant.   
The trajectory must obey the geodesic equation, 
\EQ
{d^2 x^{\mu}\over d\tau^2} + \Gamma_{\alpha \beta}^{\mu}
{d x^{\alpha}\over d\tau}{d x^{\beta} \over d\tau} = 0
\label{eq27}
\EN
with $d\tau = {ds \over \lambda(s)} $,  corresponding 
to the conservation of the energy-momentum tensor 
\EQ
D_{\mu}T^{\mu\lambda} =0
\EN
which is required by the Bianchi identity for the Einstein tensor 
$D_{\mu}G^{\mu\nu} \equiv 0$.  Since a D-particle is 
assumed to have a definite (constant) 
light-like momentum $p_-=
N/R$ in the $x_-$ direction, it is appropriate, following 
ref. \cite{bbpt},  to perform 
 the Legendre transformation
\EQ
S_p \rightarrow S_p - \int ds \, p_-{d x^-\over ds} ,
\EN
\EQ
p_-=K\lambda(s) g_{\mu -}{d x^{\mu} \over ds} .
\EN
The conventions for the light-like 
components are $x^{\pm}=x^{11}\pm t$, 
$A\cdot B = {1\over 2}(A^+B^- + A^-B^+) + A_iB_i , 
2A_-=A^+, 2A_+=A^-$.  Throughout the present paper, 
our choice of the time coordinate for D-particles is 
$\tau = x^+/2$.  

The exact solution corresponding to a D-particle 
with zero-transverse velocity is given by \cite{as}
\EQ
g_{\mu\nu} = \eta_{\mu\nu} + h_{\mu\nu}
\EN 
\EQ
h_{\mu\nu}={15\over (2\pi)^4}\kappa_{11}^2p_- \delta(x^-)
{s_{\mu}s_{\nu} \over r^7} 
\label{eq28}
\EN
\EQ
T^{\mu\nu}=p_-\delta(x^-)\delta^9(x_{\perp})s^{\mu}s^{\nu}
\EN
where $s^{\mu}$ is the velocity vector 
$s^{\mu} = {dx^{\mu} \over d\tau}$ of the trajectory 
\EQ
(s^{+}, s^-, s_{\perp}) = 2(1, 0, 0)
\label{eq214}
\EN
and $r$ is the transverse distance from the 
D-particle source. 
The exact nature of the solution 
comes from the properties that 
only non-vanishing component of the field $h_{\mu\nu}$ 
is $h_{--}$ and $\partial_+ h_{--}=0$. Then we can check that the 
Einstein equation reduces to the linearized Laplace equation 
\EQ
-{1\over 2}\triangle h_{\mu\lambda} = \kappa_{11}^2 T_{\mu\lambda}
\EN
where $\triangle$ is the Laplacian in the transverse space whose 
inverse is given as
\EQ
\langle x | \triangle^{-1} |y \rangle =-{15 \over 2(2\pi)^4 |x-y|^7} .
\EN 
 Note that both the trace and 
the divergence of the field (\ref{eq28}) vanish. 
The longitudinal momentum is 
related to the constant $K$ by 
\EQ
p_-= Kg_{+-}{dx^+ \over  d\tau}=K .
\EN
The BPS property explained by the 
linearized Laplace equation corresponds to the well known fact 
that there is no force acting between parallel light  
pencils in General Relativity. 
Because of this property, we can take an 
average over the compactified null direction 
$x^-$ of circumference $2\pi R$ to represent the 
state of definite $p_-$ momentum 
classically. The energy momentum tensor and the gravitational 
field then take the forms 
\EQ
T^{\mu\nu}={N\over 2\pi R^2}\delta^9(x_{\perp})s^{\mu}s^{\nu} ,
\label{eq217}
\EN
\EQ
h_{\mu\nu}={15\over (2\pi)^4}\kappa_{11}^2{N\over 2\pi R^2}
{s_{\mu}s_{\nu} \over r^7} .
\label{eq218}
\EN
The equation of the energy-momentum conservation 
 coincides with the ordinary flat space condition
\EQ
\partial_{\mu}T^{\mu\nu}=0 .
\EN 
It should also be noted that the gravitational 
field (\ref{eq218}) satisfies $h_{\mu\nu}s^{\mu}=0$ 
which ensures that the self-gravitating effect vanishes 
for D-particles.  This is important for the consistency 
of the interpretation of the D-particle as a singular limit of 
the gravitational wave packet in 11 dimensions.  

Generalization to D-particles with non-zero velocities 
in the transverse direction is obvious :
The velocity vector (\ref{eq214}) is replaced by 
\EQ
(s^+, s^-, s_i) = (2, -{1\over 2} v^2, v_i)
\EN
with $v_i\equiv {dx_i \over d\tau}$
and the transverse distance is now 
\EQ
r=|x_i - x_i(\tau)|^2
\EN 
with $x_i(\tau)$ being the trajectory of the D-particle with 
respect to the transverse coordinates, $x_i(\tau)=
x_i(0) + v_i\tau$.  Besides these two changes, 
the transverse $\delta$ function in the 
expression (\ref{eq217}) must be replaced by 
$\delta^9(x_i - x_i(\tau))$ which will be denoted  
by using the same notation $\delta(x_{\perp})$ below 
as in the case of zero-velocity. 

Let us now consider the system of many D-particles 
in which their relative transverse velocities 
are not zero.  In this case, simple superposition 
of the above solutions does not give exact solutions. 
However, we can solve the field equations and the 
equations of motion 
successively  by making the 
expansion with respect to the 
gravitational constant $\kappa_{11}^2$, or equivalently to the 
linearized gravitational field, and 
derive the effective action for the system of many 
D-particles by expressing the 
gravitational field in terms of the 
D-particle coordinates.   
Note that, because of the Legendre transformation, 
the action $S_D$ for D-particle 
sources is now 
\EQ
S_D = -\sum_a \int d\tau \, p_{a-} \, {d x_a^-\over d\tau} 
\label{eq222}
\EN 
Note that the original particle action $S_p$ vanishes 
owing to the massless constraint (\ref{eq26}).  
The index $a$ 
labels D-particles in the system and $p_{a-}=
N_a/R$.  

\subsection{The second order solution of the 
Einstein equation}
Our task is now to solve the Einstein equation (\ref{eq23}) and 
the D-particle equations of motion (\ref{eq27}) with the massless 
constraint (\ref{eq26}) to the next order beyond the 
linearized approximation.  

Let us begin by collecting some relevant formulas. 
The Einstein tensor to the second order with respect to 
$h_{\mu\lambda}$ is given as
\EQ
G_{\mu\lambda}= G^{(1)}_{\mu\lambda}+G^{(2)}_{\mu\lambda} ,
\EN
\EQA
G^{(1)}_{\mu\lambda}&=&
{1\over 2}(-\partial^2 h_{\mu\lambda} -\partial_{\mu}\partial_{\lambda} h^{\nu}_{\nu} 
+\partial_{\nu}\partial_{\lambda}h_{\mu}^{\nu}
+\partial_{\mu}\partial_{\nu}h^{\nu}_{\lambda}) \nonumber \\
&&-{1\over 2}\eta_{\mu\lambda}(-\partial^2 h_{\nu}^{\nu} 
+ \partial_{\alpha}\partial_{\beta}h^{\alpha\beta}) ,
\EQN
\EQA
G^{(2)}_{\mu\lambda}&=& 
-{1\over 2}h_{\nu\kappa}
(\partial^{\nu}\partial_{\lambda}h_{\mu}^{\kappa}
+\partial_{\mu}\partial^{\kappa}h_{\lambda}^{\nu}
-\partial^{\nu}\partial^{\kappa}h_{\mu\lambda}
-\partial_{\mu}\partial_{\lambda}h^{\nu\kappa})\nonumber\\
&&-{1\over 2}(\partial_{\mu}h_{\alpha\lambda}
+\partial_{\lambda}h_{\alpha\mu})
\partial_{\nu}h^{\alpha\nu} \nonumber\\
&&+{1\over 2}\partial_{\alpha}h_{\mu\lambda}
\partial_{\nu}h^{\nu\alpha}+
{1\over 2}
\partial_{\nu}h_{\alpha\lambda}
\partial^{\nu}h^{\alpha}_{\mu} 
-{1\over 2}\partial_{\alpha}h^{\nu}_{\lambda}\partial_{\nu}
h^{\alpha}_{\mu}\nonumber\\
&&+{1\over 4}\partial_{\lambda}h_{\alpha\nu}\partial_{\mu}
h^{\alpha\nu}
+{1\over 4}(\partial_{\mu}h^{\alpha}_{\lambda}
+\partial_{\lambda}h^{\alpha}_{\mu}
-\partial^{\alpha}h_{\mu\lambda})
\partial_{\alpha}h_{\nu}^{\nu}\nonumber\\
&&
-{1\over 2}h_{\mu\lambda}
(-\partial^2 h_{\nu}^{\nu} + \partial_{\alpha}\partial_{\beta}h^{\alpha\beta})-
{1\over 2}\eta_{\mu\lambda}R^{(2)} 
\EQN
where 
\EQA
R^{(2)}&=& h^{\alpha\beta}(\partial^2 h_{\alpha\beta} 
+\partial_{\alpha}\partial_{\beta}h_{\nu}^{\nu} 
-2\partial_{\gamma}\partial_{\alpha} 
h^{\gamma}_{\beta} )\nonumber\\
&&+{3\over 4}\partial_{\beta}h_{\alpha}^{\gamma}
\partial^{\beta}h^{\alpha}_{\gamma}
-{1\over 4}\partial_{\alpha}h_{\mu}^{\mu}
\partial^{\alpha}h_{\nu}^{\nu}
-\partial_{\beta}h_{\alpha}^{\beta}\partial_{\gamma}
h^{\alpha\gamma}
-{1\over 2}\partial_{\alpha}h_{\beta}^{\gamma}\partial_{\gamma}
h^{\alpha\beta} 
\EQN 
is the $O(h^2)$ part of the scalar curvature. 
 Here and 
in what follows, the Lorentz indices are contracted with respect to 
the flat space metric $\eta_{\mu\nu}$, unless 
otherwise specified.  Note that,  
with this convention, the Einstein equation is
\EQ
G^{(1)}_{\mu\lambda}+G^{(2)}_{\mu\lambda}=
\kappa_{11}^2 (T_{\mu\lambda}+ 
T^{\nu}_{\lambda}h_{\mu\nu}
+T^{\nu}_{\mu}h_{\lambda\nu}) .
\label{eq227}
\EN
The last two terms in the parenthesis of the right hand side 
arose due to the lowering 
of the tensor indices from the original definition of the 
energy-momentum tensor (\ref{eq24}). 
The conservation law of the energy-momentum tensor which 
is necessary for the consistency of the Einstein equation  is 
\EQ
\partial_{\mu}T^{\mu}_{\lambda}
+(\partial_{\mu}h_{\lambda\kappa}
-{1\over 2}\partial_{\lambda}h_{\mu\kappa})
T^{\mu\kappa} =0 
\label{eq229}
\EN
to this order. 
This is equivalent to the equation of motion 
for the D-particle 
\EQ
{d^2\over d\tau^2}x^{\alpha} + {d h^{\alpha}_{\nu}\over d\tau}
{dx^{\nu}\over d\tau} 
={1\over 2}\partial^{\alpha}h_{\mu\nu}
{dx^{\mu}\over d\tau}{dx^{\nu}\over d\tau} .
\label{eq230}
\EN
It is easy to show  that (\ref{eq229}) is a consequence of 
the equation of motion (\ref{eq230}). The converse is 
also true due to a general theorem given in \cite{geroch}. 

Now, for the source action of the 
D-particles, solving the massless constraint 
to the second order of the gravitational field, 
we find
\EQ
S_D = \sum_a {N_a\over R}\int d\tau \, 
\Bigl[{1\over 2}\Bigl({dx_a^i \over d\tau}\Bigr)^2 
+{1\over 2}
h_{\mu\nu}s_a^{\mu}s_a^{\nu}(1-h_{-\lambda}s_a^{\lambda})  
\Bigl] .
\label{eq231}
\EN
On the other hand, the structure of the 
pure gravity action is symbolically of the form 
\EQ
S_g = {1\over 2\kappa_{11}^2}
\int d^{11}x \Bigl(
{1\over 4}h{\cal D}h + {1\over 3}V_3 hhh
\Bigr)
\label{eq232}
\EN
to the cubic order where ${\cal D}$ and $V_3$ denote the 
differential kinetic operator and the 3-point vertex, respectively. 
In this notation, the Einstein equation is 
\EQ
\Bigl(-{1\over 2} {\cal D}h -V_3 hh\Bigr)^{\mu\nu}
 = \kappa_{11}^2 T^{\mu\nu} .
\label{eq233}
\EN 
Since our purpose is to derive the effective action for 
D-particles, we can use the field equation 
(\ref{eq233}) inside the action (\ref{eq232}) and obtain 
\EQ
S_g= {1\over 2\kappa_{11}^2}\int d^{11}x 
\Bigl(
{1\over 12}h{\cal D}h -{1\over 3}\kappa_{11}^2 hT
\Bigr)  .
\label{eq234}
\EN

We denote the energy-momentum tensor and 
the gravitational field in the linearized 
approximation using different notations  
as $\zeta_{\mu\lambda}$ and $\tau_{\mu\lambda}$, 
respectively.  Thus 
\EQ
\zeta_{\mu\lambda} 
=\sum_a  {15\over (2\pi)^4}\kappa_{11}^2{N_a\over 2\pi R^2}
{s_{a\mu}s_{a\lambda} \over r_a^7} ,
\label{eq235}
\EN
\EQ
\tau_{\mu\lambda}=\sum_a
{N_a\over 2\pi R^2}\delta^9(x_{a\perp})s_{a\mu}s_{a\lambda}
\label{eq236}
\EN
which satisfy 
\EQ
-{1\over 2}\partial^2 \zeta_{\mu\lambda}
=-{1\over 2}\triangle \zeta_{\mu\lambda} =\kappa^2_{11} \tau_{\mu\lambda}
\EN
and $\partial_{\mu}\zeta^{\mu\lambda}=\zeta_{\mu}^{\mu}=0$. 
The $\kappa_{11}^4$ term 
of the gravitational field is denoted by $\chi_{\mu\lambda}$ :
\EQ
h_{\mu\lambda}=\zeta_{\mu\lambda}+ \chi_{\mu\lambda} .
\EN

We have to compute the total action up to the 
order $\kappa_{11}^4$ so that the gravitational field 
is of the same order.   In order to ensure the 
consistency of the field equation to this order,  
the conservation equation 
(\ref{eq229}) demands 
\EQ
\partial_{\mu}T^{\mu}_{\lambda}
+(\partial_{\mu}\zeta_{\lambda\kappa}
-{1\over 2}\partial_{\lambda}\zeta_{\mu\kappa})
\tau^{\mu\kappa} =0 
\label{eq239}
\EN
which requires that each trajectory $x_a^{\mu}(\tau)$ 
of D-particles inside the energy-momentum tensor 
$\tau^{\mu\lambda}$  must be 
deformed to $x_a^{\mu} + \delta x_a^{\mu}$ 
such that the geodesic equations 
are satisfied.  
This is nothing but the recoil effect caused by 2-body force. 
Therefore this effect can be neglected for obtaining 
the genuine 3-body forces.    
We thus arrive at the equation which determines the 
second order term $\chi_{\mu\lambda}$ 
of the gravitational field $h_{\mu\lambda}$ in terms of the 
field $\zeta_{\mu\lambda}$ of the linear approximation :
\EQ
G^{(1)}_{\mu\lambda}(\chi) + 
G^{(2)}_{\mu\lambda}(\zeta) = \kappa_{11}^2\Bigl[
\tau_{\lambda}^{\nu}\zeta_{\mu\nu} 
+\tau_{\mu}^{\nu}\zeta_{\lambda\nu}
 +(recoil)_{\mu\lambda} \Bigr]
\label{eq243}
\EN
where the $\chi$ and $\zeta$ in the left hand side 
mean that the gravitational field $h$ is now 
$\chi$ or $\zeta$, respectively,  and $(recoil)_{\mu\lambda}$  
collectively represents the recoil effect which can be 
effectively treated as zero in obtaining the 
effective multi-body forces for D-particles. 
For the purpose of comparison with Matrix theory calculation 
in the next section where the recoil effect is 
completely neglected, this is sufficient. 
However, it should be kept in mind that 
taking into account the recoil effect is 
important for full computation of the action 
of D-particles. 

Then by substituting $h_{\mu\nu}=\zeta_{\mu\nu}+
\chi_{\mu\nu}$ in 
(\ref{eq234}) and using $({\cal D}\zeta)^{\mu\nu} = 
-2\kappa_{11}^2\tau^{\mu\nu}$, 
the gravitational action to the order 
$\kappa_{11}^4$ is given by 
\EQ
S_g = -\int d^{11}x \,  ({1\over 4}\zeta_{\mu\nu} 
\tau^{\mu\nu}
 +{1\over 3}\chi_{\mu\nu} \tau^{\mu\nu}) . 
\label{eq244}
\EN
This shows that, since the lowest energy-momentum tensor $\tau^{\mu\lambda}$ 
is traceless,  we can safely neglect the trace part of 
the second order gravitational field 
$\chi_{\mu\nu}$ within our approximation. 
Taking into account all these properties, we 
can write down the equation for the 
second order gravitational field $\chi$
\EQA
{1\over 2}\partial^2 \chi_{\mu\lambda}
&=& -{1\over 2}\zeta_{\nu\kappa}
(\partial^{\nu}\partial_{\lambda}\zeta_{\mu}^{\kappa}
+\partial_{\mu}\partial^{\kappa}\zeta^{\nu}_{\lambda}
-\partial^{\nu}\partial^{\kappa}\zeta_{\mu\lambda}
-\partial_{\mu}\partial_{\lambda}\zeta^{\nu\kappa})\nonumber\\
&& -{1\over 2}\partial_{\alpha}\zeta_{\nu\lambda}
\partial^{\nu}\zeta^{\alpha}_{\mu} +{1\over 4}\partial_{\lambda}\zeta_{\alpha\nu}\partial_{\mu}
\zeta^{\alpha\nu}\nonumber \\
&&
+{1\over 2}\partial_{\nu}\zeta_{\alpha\lambda}
\partial^{\nu}\zeta^{\alpha}_{\mu} 
  -\kappa_{11}^2(\tau_{\lambda}^{\nu}\zeta_{\mu\nu} 
+\tau_{\mu}^{\nu}\zeta_{\lambda\nu})
\nonumber \\
&& + (tracepart)_{\mu\lambda}+(recoil)_{\mu\lambda}
\label{eq245}
\EQN
where the term $(tracepart)_{\mu\lambda}$, 
denoting the part proportional to $\eta_{\mu\lambda}$, 
 does not contribute to the effective action, 
and the term $(recoil)_{\mu\lambda}$ 
can be neglected for our purpose. 
Here we have used the fact that the linearized field 
$\zeta_{\mu\nu}$ has no divergence and trace, 
and, for $\chi_{\mu\nu}$, have chosen the standard 
de Donder gauge $\partial^{\mu}\psi_{\mu\nu}\equiv \partial^{\mu}(
\chi_{\mu\nu}-{1\over 2}\eta_{\mu\nu}
\chi_{\alpha}^{\alpha})=0$ which leads to $G^{(1)}_{\mu\nu}(\chi) 
=-{1\over 2}\partial^2\psi_{\mu\nu}$ .  
This gauge choice is possible, since we have 
ensured the consistency of the Einstein equation in 
the successive approximation in the expansion 
with respect to $\kappa_{11}^2$. 
Note that the divergence of the field equation (\ref{eq243}) 
leads to the condition on the energy-momentum tensor 
which is equivalent with the conservation law 
and hence with the equation of motion. 
What we have shown above is that an apparent inconsistency 
which appears if one retains only the terms 
form the first to the third lines in the 
equation (\ref{eq245}) can be attributed to the recoil effect.  

Since the trace part does not contribute to the effective action, 
we can actually neglect the difference between 
the $\chi_{\mu\nu}$ and $\psi_{\mu\nu}=
\chi_{\mu\nu}-{1\over 2}\eta_{\mu\nu}
\chi_{\alpha}^{\alpha}$.  In what follows, for simplicity of the 
notation, we will always suppress 
those terms which do not contribute to the final 
effective action.  

Since the field $\zeta_{\mu\nu}$ is independent of $x^-$, 
the equation (\ref{eq245}) can be 
inverted using the Laplacian in the transverse space.  
We call the contributions from the first and the second lines 
in the right hand side 
the Y-contribution, and those from the 
third line the V$_g$-contribution.  Substituting the explicit forms 
for the linearized fields (\ref{eq235}) and 
 (\ref{eq236}), we find that the 
Y-contribution at the transverse position $x_a$ 
can be written as 
\EQA
\chi_{\mu\lambda}^{{\rm Y}}(x_a)  &=&-\sum_{b, c} 
{(15)^3 N_bN_c \over 4(2\pi)^4R^4M^{18}}
\Bigl[
-{1\over 2}(s_b\cdot s_c) 
(s_b\cdot \tilde{\partial}_c)
(s_{c\mu}\tilde{\partial}_{c\lambda}+ s_{c\lambda}\tilde{\partial}_{c\mu})
\nonumber \\
&&+{1\over 2}(s_b\cdot \tilde{\partial}_c)^2 s_{c\mu}s_{c\lambda} 
+{1\over 2}(s_b \cdot s_c)^2 \tilde{\partial}_{c\mu}\tilde{\partial}_{c\lambda}
\nonumber \\
&&-{1\over 2}(s_b \cdot \tilde{\partial}_c)(s_c \cdot\tilde{\partial}_b)
s_{b\lambda}s_{c\mu} +{1\over 4}(s_b\cdot s_c)^2 
\tilde{\partial}_{b\lambda}\tilde{\partial}_{c\mu}
\Bigr]\Delta(a,b,c)
\label{eq246}
\EQN
where 
\[
\Delta(a,b,c) \equiv
\int d^9 y {1\over |x_a-y|^7 |x_b -y|^7|x_c-y|^7} 
\]
\EQ
= {64(2\pi)^3 \over (15)^3}
\int_0^{\infty} d^3\sigma 
(\sigma_1\sigma_2 + \sigma_2\sigma_3 + \sigma_3\sigma_1)^{3/2}
 \exp \bigl( 
-\sigma_1|x_a-x_b|^2 -\sigma_2|x_b -x_c|^2 -\sigma_3|x_c-x_a|^2 \bigr) 
\label{eq247}
\EN
with $d^3 \sigma=d\sigma_1  d\sigma_2 d\sigma_3 $ 
and the notation $\tilde{\partial}$ is defined by 
$\tilde{\partial}_{\mu}=(\partial_+, 0, -\partial_i)$. 

For the V$_g$-contribution, we use the equality 
\EQ
\triangle^{-1} \partial_{\nu} \zeta_{\alpha\mu} 
\partial^{\nu} \zeta^{\alpha}_{\lambda} 
 =
{1\over 2}\zeta_{\alpha\lambda}\zeta_{\mu}^{\alpha}
+ \kappa_{11}^2 \triangle^{-1}
(\tau_{\alpha\lambda}\zeta_{\mu}^{\alpha} 
+\zeta_{\alpha\lambda}\tau_{\mu}^{\alpha})
\EN
for the first term in the third line  
and, after summing up all terms,  obtain  
\EQ
\chi^{V_g}_{\mu\lambda}(x_a) =\sum_{b,c}
{(15)^2 N_bN_c \over 8R^4 M^{18}} 
(s_b \cdot s_c) s_{b\lambda}s_{c\mu}
\Bigl({1\over r_{ab}^7}{1\over r_{ac}^7}
+{1\over r_{ab}^7} {1\over r_{bc}^7}
+{1\over r_{ac}^7}  
{1\over r_{bc}^7}\Bigr)
\label{eq249}
\EN
where $r_{bc}$ is the transverse distance between 
the D-particles : $r_{bc}=|x_b -x_c|$.  

\subsection{The effective action of D-particles}

We can now derive the effective action. The contribution from the pure gravity 
part  (\ref{eq244}) is   
\EQ
S_g=-{1\over 4}\sum_a {N_a\over R}\int d\tau \, \zeta_{a\mu\lambda}
s_a^{\mu} s_a^{\lambda} 
-{1\over 3}\sum_a{N_a\over R}\int d\tau \,  \chi_{a\mu\lambda}
s_a^{\mu} s_a^{\lambda} .
\EN
The source part (\ref{eq231}) gives 
\EQ
S_D = \sum_a {N_a\over R}\int d\tau \, 
\Bigl[{1\over 2}\Bigl({dx_a^i \over d\tau}\Bigr)^2 
+{1\over 2}
\zeta_{a\mu\nu}s_a^{\mu}s_a^{\nu}
+{1\over 2}
\chi_{a\mu\nu}s_a^{\mu}s_a^{\nu}
-{1\over 2}
\zeta_{a\mu\nu}\zeta_{a-\lambda}s_a^{\mu}s_a^{\nu}s_a^{\lambda}
\Bigr]  .
\EN
Thus the total effective action is 
\EQ
S_{eff}=\sum_a \int d\tau 
{N_a \over R}
\Bigl[{1\over 2}\Bigl({dx_a^i \over d\tau}\Bigr)^2 
+{1\over 4}
\zeta_{a\mu\nu}s_a^{\mu}s_a^{\nu}
+{1\over 6}
\chi_{a\mu\nu}s_a^{\mu}s_a^{\nu}
-{1\over 2}
\zeta_{a\mu\nu}\zeta_{a-\lambda}s_a^{\mu}s_a^{\nu}s_a^{\lambda}
\Bigr] .
\label{eq252}
\EN
The second term gives the familiar 2-body 
lagrangian 
\EQ
L_2 = \sum_{a<b} {15N_aN_b \over 16R^3 M^9}{v_{ab}^4 \over r_{ab}^7} .
\label{eq253}
\EN
We have used the relation $s_a\cdot s_b = -{1\over 2}(v_a-v_b)^2
\equiv -{1\over 2}v_{ab}^2$ 
with $v_a = {dx_a \over d\tau}$. 
The third and the fourth terms contain the 3-body force. 
Using (\ref{eq246}) and (\ref{eq249}), we express 
the 3-body term as a sum of two terms
\EQ
L_3 = L_V +L_Y
\EN
Note that the 
V-type contribution consists of two parts 
corresponding to the contribution 
 of (\ref{eq249}) and the last term of  (\ref{eq252}) 
\EQA
L_V&=& \sum_{a,b,c}{(15)^2 N_aN_bN_c\over 48R^5M^{18}}
(s_b\cdot s_c)(s_a\cdot s_b)(s_a\cdot s_c) 
\Bigl(
{1\over r_{ab}^7}{1\over r_{ac}^7}
 +
 {1\over r_{ab}^7} {1\over r_{bc}^7}
+{1\over r_{ac}^7}  
{1\over r_{bc}^7}\Bigr)\nonumber\\
&&-
\sum_{a,b,c}{(15)^2 N_aN_bN_c\over 8R^5M^{18}}
(s_b\cdot s_a)^2(s_c\cdot s_a){1\over r_{ab}^7}{1\over r_{ca}^7} .
\EQN
In terms of the relative velocities, the sum of two terms 
 is rewritten as 
\EQ
L_V=  -\sum_{a,b,c}{(15)^2 N_aN_bN_c\over 64 R^5M^{18}}
v_{ab}^2 v_{ca}^2 (v_{ca}\cdot v_{ab})
{1\over r_{ab}^7}{1\over  r_{ca}^7} .
\label{eq256}
\EN
The Y contribution is 
\EQA
L_Y  &=&-\sum_{a, b, c} 
{(15)^3 N_aN_bN_c \over 24(2\pi)^4R^5M^{18}}
\Bigl[
-(s_b\cdot s_c) 
(s_{c}\cdot s_a)(s_b\cdot \tilde{\partial}_c)
(s_a\cdot \tilde{\partial}_{c})\nonumber \\
&&+{1\over 2}( s_{c}\cdot s_a)^2(s_b\cdot \tilde{\partial}_c)^2  
+{1\over 2}(s_b \cdot s_c)^2 (s_a\cdot \tilde{\partial}_{c})^2
\nonumber \\
&&-{1\over 2}(s_{b}\cdot s_a)
(s_a\cdot s_{c})(s_b \cdot \tilde{\partial}_c)(s_c \cdot\tilde{\partial}_b)
 \nonumber\\
&&+{1\over 4}(s_b\cdot s_c)^2 
(s_a\cdot \tilde{\partial}_{b})(s_a\cdot \tilde{\partial}_{c})
\Bigr]\Delta(a,b,c) .
\EQN
Using $s_a\cdot \tilde{\partial}_c = {\partial\over \partial \tau_c} 
- v_a\cdot \nabla_c
=v_{ca}\cdot \nabla_c $ which is valid because the 
second derivatives with respect to time can be neglected 
in the present approximation, 
we can rewrite this as 
\EQA
L_Y  &=&-\sum_{a, b, c} 
{(15)^3 N_aN_bN_c \over 96(2\pi)^4R^5M^{18}} 
\Bigl[
-v_{bc}^2v_{ca}^2(v_{cb}\cdot \nabla_c)
(v_{ca}\cdot \nabla_c)\nonumber \\
&&+{1\over 2} v_{ca}^4(v_{cb}\cdot \nabla_c)^2  
+{1\over 2}v_{b c}^4 (v_{ca}\cdot \nabla_c)^2
\nonumber \\
&&-{1\over 2}v_{ba}^2
v_{ac}^2(v_{cb}\cdot \nabla_c)(v_{bc}\cdot \nabla_b)
 \nonumber\\
&&+{1\over 4}v_{bc}^4 
( v_{ba}\cdot \nabla_b)(v_{ca}\cdot \nabla_c)
\Bigr]\Delta(a,b,c) .
\label{eq258}
\EQN
The results (\ref{eq256}) and (\ref{eq258}) are our 
final form of the 3-body interaction Lagrangian of D-particles. 

Before proceeding to the investigation of the corresponding 
matrix model results, it is appropriate here to make some 
comments on the properties of the above effective Lagrangian. 

First, the Y-type interaction (\ref{eq258}) vanishes when any 
pair of two D-particles are parallel with vanishing relative 
velocities.  Furthermore, only remaining V-type interaction 
for such cases is 
\EQ
2{(15)^2 N_aN_bN_c\over 64 R^5M^{18}}
v_{ab}^6 {1\over r_{ab}^7}{1\over  r_{ca}^7} 
\EN
where the parallel particles are $b$ and $c$. 
This is just as expected from the BPS property of the 
D-particles, since the gravitational fields 
produced by parallel D-particles 
are exactly the superposition of the linearized fields 
produced by them.  

Secondly, it is remarkable that the kinematical 
factor of the V-type contributions is 
reduced to the special combination
\[
v_{ab}^2v_{ca}^2 (v_{ca} \cdot v_{ab})
={1\over 2}v_{ab}^2v_{ca}^2
(v_{bc}^2-v_{ca}^2 -v_{ab}^2).
\]
   From the viewpoint of Matrix theory,  
this form with the given 
$r$ dependence is just the  typical kinematical 
factor which arises from  a 6-point two-loop diagram 
of `$\infty$' type 
in the `quasi-static' approximation (see, e.g. \cite{kabat}), 
 with one 4-point vertex $\Tr[A,Y]^2$ and 
three external lines in each loop, 
entering through the 2-point vertices $\Tr(A[\dot{B},Y])$,   
where $A$ and $Y$  
are the off-diagonal parts of 
 the gauge field and Higgs fields, respectively.  
The $\dot{B}$ is the time derivative of the diagonal background 
$B$ of the Higgs field, 
corresponding to the velocity of  D-particles. 
This suggests that Matrix theory 
can be consistent with supergravity, contrary 
to the claim of ref. \cite{dine}.  
Then the question is whether the other possible types of 
kinematical factors are canceled and whether the 
coefficients and also the Y-type contribution 
come out correctly.  In the next section, 
we will confirm this expectation by carrying out a 
complete calculation of the two-loop diagrams without 
relying upon the quasi-static approximation. 

Thirdly, it should be noted that in the derivation 
of the 2-body interaction in ref. \cite{bbpt}  only the 
 action for the probe D-particle 
is taken into account. In contrast with 
this, in our derivation, the pure gravity action 
is included.  We emphasize that, for the derivation 
of the effective action for general configurations  
of D-particles in which all the D-particles are 
treated democratically as in our formulation, 
it is essential to include the pure gravity action. 
In the system of two D-particles, we can always take 
the Galilei frame such that one of them (source) is at rest 
in the transverse space. In this case, the total action excluding the 3(and higher)-body interactions can be reduced to the 
single particle action for the probe since 
the pure gravity action just cancels the 
action of the source.   

Finally,  in our derivation of the effective action 
we have represented D-particles as the 
singular solution of the Einstein equation. 
Correspondingly, we have introduced the source action 
for D-particles. However, D-particles are 
supposed to be the special states of the 11 dimensional graviton 
super multiplet.  In this sense, the point-like source 
variables in the lowest energy-momentum tensor 
$\tau_{\mu\nu}$  should be interpreted as 
the collective coordinate of the background 
gravitational field. From this viewpoint, a conceptually 
more satisfactory treatment would be to introduce some 
non-singular D-particle background with finite extension 
for the metric and take the point-like limit  afterwards. 
In the present paper, however, we do not elaborate along this line 
since we expect that both treatments lead to the same results.  
For examples of such treatments in the case of a 
test point particle in General 
Relativity, see ref. \cite{infeld} and references therein.

 \section{Multi-body D-particle scattering  in Matrix theory}
\setcounter{equation}{0}
Let us now discuss the multi-body 
scattering of D-particles or Kaluza-Klein 
gravitons in Matrix theory up to the two-loop approximation. 
To make the present paper reasonably self-contained,  
we start from summarizing some basic facts and formulas 
of Matrix theory. 
 
\subsection{The setup}
Matrix theory is defined by the action of
the supersymmetric Yang-Mills theory
dimensionally reduced from 9+1 dimensions to 0+1 dimension
\begin{eqnarray}
S = \int dt \left[ \frac{\kappa}{2} {\rm tr} D_t X^n D_t X^n 
+ \frac{\kappa}{4} g^2 {\rm tr} [X^n,X^m] [X^n,X^m] \right. 
\nonumber \\
\left. + \frac{\kappa}{2} {\rm tr}
( i \theta^T D_t \theta
+ g \theta^T \gamma^n [X^n,\theta]) \right],
\end{eqnarray}
where $g$ is the Yang-Mills coupling constant and
$n,m = 1,2,\cdots,9$ stand for transverse dimensions.
The covariant derivative is defined by
\begin{eqnarray}
D_t X^n = \partial_t X^n -ig [A,X^n], \quad
D_t \theta = \partial_t \theta -ig [A,\theta].
\end{eqnarray}
$\gamma^n$ are $SO(9)$ gamma matrices in Majorana representation.
We take $\gamma^n$ to be real and symmetric matrices satisfying
$\{ \gamma^n,\gamma^m \} = 2\delta^{nm}$
and $\theta$ to be real.
$X^n$, $A$ and $\theta$ are hermitian $U(N)$ matrices. 
For the Grassmann matrix $\theta$, the reality condition 
is defined by the Majorana condition. 
The parameters $\kappa$ and $g$ are related to the 
M-theory parameters as 
$\kappa=1/R, g=M^3 R$. 

We will perform our computations in Euclidean formulation, 
defining the Euclidean time $\tau$ and gauge field in Euclidean time
$\tilde{A}$ as
\begin{eqnarray}
\tau = it, \quad \tilde{A} = -iA. 
\end{eqnarray}
The time variable $\tau$ of the previous section should be 
identified with $t $. 
The Euclidean action is then defined by
\begin{eqnarray}
S_E = \int d \tau \left[ \frac{\kappa}{2} {\rm tr} D_\tau X^n D_\tau X^n 
- \frac{\kappa}{4} g^2 {\rm tr} [X^n,X^m] [X^n,X^m] \right. 
\nonumber \\
\left. + \frac{\kappa}{2} {\rm tr}
( \theta^T D_\tau \theta
- g \theta^T \gamma^n [X^n,\theta]) \right],
\end{eqnarray}
where the definition of the covariant derivative has changed to
\begin{eqnarray}
D_\tau X^n = \partial_\tau X^n -ig [\tilde{A},X^n], \quad
D_\tau \theta = \partial_\tau \theta -ig [\tilde{A},\theta].
\end{eqnarray}
We expand the action around the classical background field $B^n$
as
\begin{eqnarray}
X^n = \frac{1}{g} B^n + Y^n.
\end{eqnarray}
We choose the background configuration to be 
the straight-line trajectories of $N$ D-particles
\begin{eqnarray}
B^n_{ij} = \delta_{ij} (x^n_i + \tilde{v}^n_i \tau)
\label{Bdef}
\end{eqnarray}
where $i,j$ are $U(N)$ indices.  Note that the 
background matrices are actually assumed to 
consist of several block submatrices 
which are proportional to the  unit matrices and that 
the order $N_a \, \, (a=1, 2, \ldots) (N=\sum_a N_a) $ of each block 
determines the mass $N_a/R$ of the  
corresponding D-particle. 
Note also that we have rescaled the 
background coordinates at time $\tau=0$ and velocities 
by $x \rightarrow {1\over g}x, v\rightarrow {i\over g} \tilde{v}$. 
Throughout the present section, the following
abbreviations  will be used:
\begin{eqnarray}
&& r^n_i(\tau) = x^n_i + \tilde{v}^n_i \tau, 
\nonumber \\
&& r^n_{ij}(\tau) = r^n_i(\tau) -r^n_j(\tau), \quad
r_{ij}(\tau)^2 
= r^n_{ij}(\tau) r^n_{ij}(\tau), \quad 
r_{ij}(\tau) = \sqrt{r_{ij}(\tau)^2}, \quad
r\!\!\!/_{ij}(\tau) = \gamma^n r^n_{ij}(\tau), 
\nonumber \\
&& x^n_{ij} = x^n_i -x^n_j, \quad
x_{ij}^2 = x^n_{ij} x^n_{ij}, \quad x_{ij} = \sqrt{x_{ij}^2}, \quad
x\!\!\!/_{ij} = \gamma^n x^n_{ij}, 
\nonumber \\
&& \tilde{v}^n_{ij} = \tilde{v}^n_i -\tilde{v}^n_j, \quad
\tilde{v}_{ij}^2 = \tilde{v}^n_{ij} \tilde{v}^n_{ij}, \quad \tilde{v}_{ij} = \sqrt{\tilde{v}_{ij}^2}, \quad
\tilde{v}\!\!\!/_{ij} = \gamma^n \tilde{v}^n_{ij} ,\quad 
x_{ij} \cdot \tilde{v}_{ij} = x^n_{ij} \tilde{v}^n_{ij}.
\end{eqnarray}
We use the standard  background field gauge condition
\begin{eqnarray}
-\partial_\tau \tilde{A} +i[B^n,Y^n] = 0,
\end{eqnarray}
with  the gauge-fixed action  
\begin{eqnarray}
\tilde{S} &=& \int d \tau \left[ \right.
\frac{\kappa}{g^2} {\rm tr}~ 
\frac12 \partial_\tau B^n \partial_\tau B^n 
\nonumber \\ && \qquad
+ \kappa {\rm tr}~ ( \frac12 \partial_\tau Y^n \partial_\tau Y^n 
- \frac12 [B^m,Y^n][B^m,Y^n] 
+ \frac12 \partial_\tau \tilde{A} \partial_\tau \tilde{A} 
- \frac12 [B^m,\tilde{A}][B^m,\tilde{A}] 
\nonumber \\ && \qquad
- 2i \partial_\tau B^n [\tilde{A},Y^n]
- \bar{c} \partial_\tau^2 c + \bar{c} [B^m, [B^m, c]]
+\frac{1}{2} \theta^T 
\partial_\tau \theta
- \frac{1}{2} \theta^T \gamma^n [B^n,\theta] ) 
\nonumber \\ && \qquad
+ g \kappa~ {\rm tr}~ (-i\partial_\tau Y^n [\tilde{A},Y^n] 
- [B^n,Y^m][Y^n,Y^m] -[B^n,\tilde{A}][Y^n,\tilde{A}] 
\nonumber \\ && \qquad
-i \partial_\tau \bar{c} [\tilde{A},c] -[B^n,\bar{c}][Y^n,c]
- \frac{i}{2} \theta^T [\tilde{A},\theta] 
- \frac{1}{2} \theta^T \gamma^n [Y^n, \theta] ) 
\nonumber \\ && \qquad
+g^2 \kappa {\rm tr}~ ( -\frac14 [Y^n,Y^m][Y^n,Y^m]
\left. -\frac12 [\tilde{A},Y^m][\tilde{A},Y^m] ) \right].
\end{eqnarray}
Now from the quadratic part of the action
\begin{eqnarray}
\tilde{S}_{(2)} 
= \int d \tau \left[ \right.
 \frac{\kappa}{2} ( \partial_\tau Y^n_{ij} \partial_\tau Y^n_{ji} 
+r_{ij}(\tau)^2 Y^n_{ij} Y^n_{ji}) 
+\frac{\kappa}{2} 
( \partial_\tau \tilde{A}_{ij} \partial_\tau \tilde{A}_{ji} 
+ r_{ij}(\tau)^2 \tilde{A}_{ij} \tilde{A}_{ji})
\nonumber \\
-2i \kappa \tilde{v}^n_{ij} \tilde{A}_{ij} Y^n_{ji}
+\kappa (-\bar{c}_{ij} \partial_\tau^2 c_{ji} 
+ r_{ij}(\tau)^2  \bar{c}_{ij} c_{ji} ) 
\nonumber \\
+ \frac{\kappa}{2} 
( \theta^T_{ij} \partial_\tau \theta_{ji}
+ \theta^{T}_{ij} r\!\!\!/_{ij}(\tau) \theta_{ji} ) 
\left. \right],
\end{eqnarray}
 we derive the propagators in proper-time representations. 

Propagators of the bosonic fields:
\begin{eqnarray}
\langle Y^n_{ij} (\tau_1) Y^m_{kl} (\tau_2) \rangle_0 
&=& \frac{\delta_{il} \delta_{jk}}
{\kappa} \int_0^\infty d\sigma~ \left(
\delta^{nm} + 2 V_{ij}^n (\sigma) V_{ij}^m (\sigma) 
\right) \Delta_{ij} (\sigma,\tau_1,\tau_2), \\
\langle \tilde{A}_{ij} (\tau_1) \tilde{A}_{kl} (\tau_2) \rangle_0
&=& \frac{\delta_{il} \delta_{jk}}
{\kappa} \int_0^\infty d\sigma~
\left(1 + 2 V_{ij} (\sigma)^2 \right) 
\Delta_{ij} (\sigma,\tau_1,\tau_2), \\
\langle Y^n_{ij} (\tau_1) \tilde{A}_{kl} (\tau_2) \rangle_0 
&=& \frac{2i \delta_{il} \delta_{jk} }
{\kappa} \int_0^\infty d\sigma~
V_{ij}^n (\sigma) C_{ij} (\sigma) \Delta_{ij} (\sigma,\tau_1,\tau_2), \\
\langle \tilde{A}_{ij} (\tau_1) Y^m_{kl} (\tau_2) \rangle_0 
&=& - \frac{2i \delta_{il} \delta_{jk} }
{\kappa} \int_0^\infty d\sigma~
V_{ij}^m (\sigma) C_{ij} (\sigma) \Delta_{ij} (\sigma,\tau_1,\tau_2), \\
\langle \bar{c}_{ij} (\tau_1) c_{kl} (\tau_2) \rangle_0
&=& \frac{ \delta_{il} \delta_{jk} }
{\kappa} \int_0^\infty d\sigma~ \Delta_{ij} (\sigma,\tau_1,\tau_2), \\
\langle c_{ij} (\tau_1) \bar{c}_{kl} (\tau_2) \rangle_0
&=& - \frac{ \delta_{il} \delta_{jk} }
{\kappa} \int_0^\infty d\sigma~ \Delta_{ij} (\sigma,\tau_1,\tau_2).
\end{eqnarray}

Propagators of the fermionic fields: 
\begin{eqnarray}
&& \langle \theta^\alpha_{ij} (\tau_1) \theta^\beta_{kl} (\tau_2) 
\rangle_0
\nonumber \\
&& = \frac{ \delta_{il} \delta_{jk} }
{\kappa} \int_0^\infty d\sigma~ \left[
- \left( 
{\bf 1} C_{ij} (\sigma) \partial_{\tau_1} 
+ r\!\!\!/_{ij} (\tau_1) V\!\!\!\!/_{ij} (\sigma)
\right)
- \left(
V\!\!\!\!/_{ij} (\sigma) \partial_{\tau_1} 
+ C_{ij} (\sigma) r\!\!\!/_{ij} (\tau_1)
\right)
\right]^{\alpha \beta}
\nonumber \\
&& \qquad \qquad \qquad \qquad \times
\Delta_{ij} (\sigma,\tau_1,\tau_2), 
\end{eqnarray}
or
\begin{eqnarray}
&& \langle \theta^\beta_{kl} (\tau_2) \theta^\alpha_{ij} (\tau_1) 
\rangle_0
\nonumber \\
&& = \frac{ \delta_{il} \delta_{jk} }
{\kappa} \int_0^\infty d\sigma~ \left[
- \left( 
{\bf 1} C_{ij} (\sigma) \partial_{\tau_2} 
+ r\!\!\!/_{ij} (\tau_2) V\!\!\!\!/_{ij} (\sigma)
\right)
+ \left(
V\!\!\!\!/_{ij} (\sigma) \partial_{\tau_2} 
+ C_{ij} (\sigma) r\!\!\!/_{ij} (\tau_2)
\right)
\right]^{\beta \alpha}
\nonumber \\
&& \qquad \qquad \qquad \qquad \times
\Delta_{ij} (\sigma,\tau_1,\tau_2).
\end{eqnarray}
Here we have defined 
\begin{eqnarray}
\Delta_{ij} (\sigma,\tau_1,\tau_2) & \equiv &
\exp \left[ -\sigma  (-\partial_{\tau_1}^2 + r^2(\tau_1) )
\right] \delta(\tau_1 -\tau_2) \nonumber \\
&=& \sqrt{\frac{\tilde{v}_{ij}}{2\pi \sinh (2\sigma \tilde{v}_{ij})}}
\exp \left[
-\frac{\tilde{v}_{ij}}{2 \sinh (2\sigma \tilde{v}_{ij})}
\left(
(\tau_1^2+\tau_2^2) \cosh (2\sigma \tilde{v}_{ij}) -2 \tau_1 \tau_2
\right) \right. \nonumber \\
& & \left.
-\frac{x_{ij} \cdot \tilde{v}_{ij} \tanh (\sigma \tilde{v}_{ij})}{\tilde{v}_{ij}}
\left( \tau_1+\tau_2+\frac{x_{ij} \cdot \tilde{v}_{ij}}{\tilde{v}_{ij}^2} \right)
-\sigma \left( x_{ij}^2 -\frac{(x_{ij} \cdot \tilde{v}_{ij})^2}{\tilde{v}_{ij}^2} 
\right)
\right] \nonumber \\
&=& \sqrt{\frac{\tilde{v}_{ij}}{2\pi \sinh (2\sigma \tilde{v}_{ij})}}
\exp \left[
-\tilde{v}_{ij} \left( \frac{\tau_1-\tau_2}{2} \right)^2
\coth (\sigma \tilde{v}_{ij})
\right. \nonumber \\ 
& & \left.
-\tilde{v}_{ij} 
\left( \frac{\tau_1+\tau_2}{2} 
+\frac{x_{ij} \cdot \tilde{v}_{ij}}{\tilde{v}_{ij}^2} \right)^2 
\tanh (\sigma \tilde{v}_{ij})
-\sigma \left( x_{ij}^2 -\frac{(x_{ij} \cdot \tilde{v}_{ij})^2}{\tilde{v}_{ij}^2} 
\right)
\right],
\label{Delta}
\end{eqnarray}
\begin{eqnarray}
V_{ij}^n (\sigma) &\equiv& 
\frac{\tilde{v}_{ij}^n}{\tilde{v}_{ij}}
\sinh (\sigma \tilde{v}_{ij}), \\
C_{ij} (\sigma) &\equiv& \cosh (\sigma \tilde{v}_{ij}), \\
V_{ij} (\sigma)^2 &\equiv& V_{ij} (\sigma) \cdot V_{ij} (\sigma) 
= V_{ij}^n (\sigma) V_{ij}^n (\sigma)
= \sinh^2 (\sigma \tilde{v}_{ij}), \\
V_{ij} (\sigma) &\equiv& \sqrt{V_{ij} (\sigma) \cdot V_{ij} (\sigma)}
= \sqrt{V_{ij}^n (\sigma) V_{ij}^n (\sigma)}
= \sinh (\sigma \tilde{v}_{ij}).
\end{eqnarray}
\subsection{The calculations of the effective action}
We now present the complete form of 
 the effective action $\tilde{\Gamma}$
around the background configuration
in the two-loop approximation
\begin{eqnarray}
\tilde{\Gamma} = \tilde{\Gamma}_{(0)} + \tilde{\Gamma}_{(1)} + \tilde{\Gamma}_{(2)} + O(g^4) 
\end{eqnarray}
where $\tilde{\Gamma}_{(\ell)}$ stands for the $\ell$-loop contribution.
 
We note that the effective action from the 
viewpoint of the Yang-Mills field theory is 
nothing but the one-particle irreducible 
contribution to the eikonal phase shift 
of D-particle, since the results contains the 
time integration of the genuine light-cone lagrangian
for straight line trajectories. 
First, the classical action  $\tilde{\Gamma}_{(0)}$ is
\begin{eqnarray}
\tilde{\Gamma}_{(0)} =
\int_{-\infty}^\infty d \tau
\frac{\kappa}{g^2} {\rm tr}~ \frac{1}{2}
\partial_\tau B^n \partial_\tau B^n
= \int_{-\infty}^\infty d \tau
\frac{1}{Rg^2} \sum_{i}
\frac{1}{2} \tilde{v}_i^2 =\int_{-\infty}^\infty d \tau
\sum_a{N_a\over Rg^2}\tilde{v}_a^2 
\end{eqnarray}
where in the last equality, we switched from the summation 
over the U($N$) indices to the block indices 
specifying the system of coincident D-particles. 

The one-loop contribution is obtained from 
the functional determinant as
\begin{eqnarray}
&& \exp [ -\tilde{\Gamma}_{(1)} ] 
= \int DY^n D \tilde{A} D \bar{c} Dc D \theta~ e^{-\tilde{S}_{(2)}}
\nonumber \\
&& = \prod_{i<j} 
{\rm det}^{-6}(-\partial_{\tau}^2 + r_{ij}(\tau)^2 )
{\rm det}^{-1}(-\partial_{\tau}^2 + r_{ij}(\tau)^2 +2 \tilde{v}_{ij} )
{\rm det}^{-1}(-\partial_{\tau}^2 + r_{ij}(\tau)^2 -2 \tilde{v}_{ij} )
\nonumber \\
&& \qquad \qquad \times
{\rm det}^{4}(-\partial_{\tau}^2 + r_{ij}(\tau)^2 + \tilde{v}_{ij} )
{\rm det}^{4}(-\partial_{\tau}^2 + r_{ij}(\tau)^2 - \tilde{v}_{ij} ).
\end{eqnarray}
$\tilde{\Gamma}_{(1)}$ is expressed in the proper-time representation as
\begin{eqnarray}
\tilde{\Gamma}_{(1)} 
&=& -\sum_{i<j} \int_{-\infty}^\infty d \tau
\int_0^\infty \frac{d\sigma}{\sigma} ~
16 \sinh^4 \frac{\sigma \tilde{v}_{ij} }{2} ~
\Delta_{ij} (\sigma,\tau,\tau) \\
&=& -\sum_{i<j} \int_0^\infty \frac{d\sigma}{\sigma} ~
16 \sinh^4 \frac{\sigma \tilde{v}_{ij} }{2} ~
\frac{1}{2 \sinh ( \sigma \tilde{v}_{ij} )}
\exp \left[ -\sigma \left(
x_{ij}^2 - \frac{(x_{ij} \cdot \tilde{v}_{ij} )^2}{\tilde{v}_{ij}^2}
\right) \right].
\nonumber \\
\label{Gamma1}
\end{eqnarray}
Note that we have slightly generalized the well-known 
result of the 2-body
scattering because we cannot, in general, assume 
vanishing of $x_{ij} \cdot \tilde{v}_{ij}$
  in multi-body scattering.
The leading contribution of 
the one-loop effective action $\tilde{\Gamma}_{(1)}$ is
\begin{eqnarray}
\tilde{\Gamma}_{(1) ~{\rm leading}}
&=& - \sum_{i<j} \int_{-\infty}^\infty d \tau
\int_0^\infty \frac{d \sigma}{\sigma}
\frac{\sigma^4 \tilde{v}_{ij}^4}{\sqrt{4 \pi \sigma}}
\exp [-\sigma (\tilde{v}_{ij} \tau +x_{ij})^2]
\nonumber \\
&=& - \int_{-\infty}^\infty d \tau \sum_{i<j}
\frac{15}{16} 
\frac{\tilde{v}_{ij}^4}
{[(\tilde{v}_{ij} \tau + x_{ij})^2]^{7/2}},
\end{eqnarray}
where we used the proper-time representation
\begin{eqnarray}
\frac{1}{[(\tilde{v} \tau +x)^2]^{7/2}} =
\frac{8}{15 \sqrt{\pi}}
\int_0^\infty d \sigma \frac{\sigma^3}{\sqrt{\sigma}}
\exp[- \sigma (\tilde{v} \tau +x)^2].
\label{r7}
\end{eqnarray}
Since the one-loop contribution has been studied 
in many works, no further discussion is needed here. 

Two loop calculations have been performed in ref. \cite{bb} 
\cite{bbpt}.   
In what follows, we will report  an extension 
of their results  in order to make possible the comparison 
of the 3-body interaction 
of D-particles  with the results of the preceding section. 
To obtain the two-loop contribution $\tilde{\Gamma}_{(2)}$
to the effective action, we have to evaluate
all one-particle irreducible (1PI) diagrams.
Since the nonplanar 1PI diagrams at two-loop order 
have only one index loop and hence do not contribute
to D-particle interactions, we can restrict our consideration to 
planar diagrams.
There are fourteen different contributions depending on the 
combinations of the propagating fields: 
\begin{eqnarray}
&& -\tilde{\Gamma}_{(2)} 
\nonumber \\
&&= \frac14 g^2 \kappa \int d\tau 
\langle {\rm tr} [Y^n(\tau),Y^m(\tau)] 
[Y^n(\tau),Y^m(\tau)] \rangle_{1PI,{\rm planar}} 
\nonumber \\
&& + \frac12 g^2 \kappa \int d\tau 
\langle {\rm tr} [\tilde{A}(\tau),Y^m(\tau)] 
[\tilde{A}(\tau),Y^m(\tau)] \rangle_{1PI,{\rm planar}} 
\nonumber \\
&& -\frac12 g^2 \kappa^2 \int d\tau_1 d\tau_2 
\langle {\rm tr} \partial_{\tau_1} Y^p(\tau_1)[\tilde{A}(\tau_1),Y^p(\tau_1)] 
~{\rm tr} \partial_{\tau_2} Y^q(\tau_2)[\tilde{A}(\tau_2),Y^q(\tau_2)] 
\rangle_{1PI,{\rm planar}} 
\nonumber \\
&& + \frac12 g^2 \kappa^2 \int d\tau_1 d\tau_2 
\langle {\rm tr} [B^n,Y^p](\tau_1)[Y^n(\tau_1),Y^p(\tau_1)] 
~{\rm tr} [B^m,Y^q](\tau_2)[Y^m(\tau_2),Y^q(\tau_2)] 
\rangle_{1PI,{\rm planar}} 
\nonumber \\
&& + \frac12 g^2 \kappa^2 \int d\tau_1 d\tau_2 
\langle {\rm tr} [B^n,\tilde{A}](\tau_1)[Y^n(\tau_1),\tilde{A}(\tau_1)] 
~{\rm tr} [B^m,\tilde{A}](\tau_2)[Y^m(\tau_2),\tilde{A}(\tau_2)] 
\rangle_{1PI,{\rm planar}} 
\nonumber \\
&& + ig^2 \kappa^2 \int d\tau_1 d\tau_2 
\langle {\rm tr} \partial_{\tau_1} Y^p(\tau_1)[\tilde{A}(\tau_1),Y^p(\tau_1)] 
~{\rm tr} [B^m,Y^q](\tau_2)[Y^m(\tau_2),Y^q(\tau_2)] 
\rangle_{1PI,{\rm planar}} 
\nonumber \\
&& + ig^2 \kappa^2 \int d\tau_1 d\tau_2 
\langle {\rm tr} \partial_{\tau_1} Y^p(\tau_1)[\tilde{A}(\tau_1),Y^p(\tau_1)] 
~{\rm tr} [B^m,\tilde{A}](\tau_2)[Y^m(\tau_2),\tilde{A}(\tau_2)] 
\rangle_{1PI,{\rm planar}} 
\nonumber \\
&& + g^2 \kappa^2 \int d\tau_1 d\tau_2 
\langle {\rm tr} [B^n,Y^p](\tau_1)[Y^n(\tau_1),Y^p(\tau_1)] 
~{\rm tr} [B^m,\tilde{A}](\tau_2)[Y^m(\tau_2),\tilde{A}(\tau_2)] 
\rangle_{1PI,{\rm planar}} 
\nonumber \\
&& - \frac12 g^2 \kappa^2 \int d\tau_1 d\tau_2 
\langle {\rm tr} \partial_{\tau_1} \bar{c}(\tau_1)
[\tilde{A}(\tau_1),c(\tau_1)] 
~{\rm tr} \partial_{\tau_2} \bar{c}(\tau_2)
[\tilde{A}(\tau_2),c(\tau_2)] \rangle_{1PI,{\rm planar}} 
\nonumber \\
&& + \frac12 g^2 \kappa^2 \int d\tau_1 d\tau_2 
\langle {\rm tr} [B^n,\bar{c}](\tau_1)[Y^n(\tau_1),c(\tau_1)] 
~{\rm tr} [B^m,\bar{c}](\tau_2)[Y^m(\tau_2),c(\tau_2)] 
\rangle_{1PI,{\rm planar}} 
\nonumber \\
&& + ig^2 \kappa^2 \int d\tau_1 d\tau_2 
\langle {\rm tr} \partial_{\tau_1} \bar{c}(\tau_1)
[\tilde{A}(\tau_1),c(\tau_1)] 
~{\rm tr} [B^m,\bar{c}](\tau_2)
[Y^m(\tau_2),c(\tau_2)] \rangle_{1PI,{\rm planar}} 
\nonumber \\
&& - \frac18 g^2 \kappa^2 \int d\tau_1 d\tau_2 
\langle {\rm tr} \theta^\alpha(\tau_1)
[\tilde{A}(\tau_1),\theta^\alpha(\tau_1)] 
~{\rm tr} \theta^\beta(\tau_2)
[\tilde{A}(\tau_2),\theta^\beta(\tau_2)] \rangle_{1PI,{\rm planar}} 
\nonumber \\
&& + \frac18 g^2 \kappa^2 
\gamma^n_{\alpha \beta} \gamma^m_{\gamma \delta}
\int d\tau_1 d\tau_2 
\langle {\rm tr} \theta^\alpha(\tau_1)
[Y^n(\tau_1),\theta^\beta(\tau_1)] 
~{\rm tr} \theta^\gamma(\tau_2)
[Y^m(\tau_2),\theta^\delta(\tau_2)] \rangle_{1PI,{\rm planar}} 
\nonumber \\
&& +\frac{i}{4} g^2 \kappa^2 
\gamma^n_{\beta \gamma}
\int d\tau_1 d\tau_2 
\langle {\rm tr} \theta^\alpha(\tau_1)
[\tilde{A}(\tau_1),\theta^\alpha(\tau_1)] 
~{\rm tr} \theta^\beta(\tau_2)
[Y^n(\tau_2),\theta^\gamma(\tau_2)] \rangle_{1PI,{\rm planar}}.
\nonumber \\
\label{Gamma2}
\end{eqnarray}
Evaluating these terms is a straightforward but extremely tedious task. 
After substituting the explicit forms of the propagators in 
the proper time representation, 
we can arrange the result in the following way:
\begin{eqnarray}
&& -\tilde{\Gamma}_{(2)} = 
\frac{g^2}{\kappa} \int d \tau_1 d \tau_2 d\sigma_1 d\sigma_2 d\sigma_3
\sum_{i,j,k} \left[ \right.
\nonumber \\
&& \qquad \qquad
~~ P_q (\{ \sigma \},\{ \tilde{v} \} )
 \{ -\partial_{\sigma_1} \Delta_{ij} (\sigma_1,\tau_1,\tau_2) \}
\Delta_{jk} (\sigma_2,\tau_1,\tau_2) 
\Delta_{ki} (\sigma_3,\tau_1,\tau_2) 
\nonumber \\
&& \qquad \qquad 
+ P_1 (\{ \sigma \},\{ \tilde{v} \} )
\{ \partial_{\tau_1} \partial_{\tau_2}
\Delta_{ij} (\sigma_1,\tau_1,\tau_2) \}
\Delta_{jk} (\sigma_2,\tau_1,\tau_2) 
\Delta_{ki} (\sigma_3,\tau_1,\tau_2) 
\nonumber \\
&& \qquad \qquad 
+ P_2 (\{ \sigma \},\{ \tilde{v} \},\{ \tilde{v} \cdot r(\tau_2) \} )
\{ \partial_{\tau_1}
\Delta_{ij} (\sigma_1,\tau_1,\tau_2) \}
\Delta_{jk} (\sigma_2,\tau_1,\tau_2) 
\Delta_{ki} (\sigma_3,\tau_1,\tau_2) 
\nonumber \\
&& \qquad \qquad 
+ P_3 (\{ \sigma \},\{ \tilde{v} \} )
( r_{ij}(\tau_1) \cdot r_{ij}(\tau_2) )
\Delta_{ij} (\sigma_1,\tau_1,\tau_2) 
\Delta_{jk} (\sigma_2,\tau_1,\tau_2) 
\Delta_{ki} (\sigma_3,\tau_1,\tau_2) 
\nonumber \\
&& \qquad \qquad 
+ P_4 (\{ \sigma \},\{ \tilde{v} \},
\{ \tilde{v} \cdot r(\tau_1) \},\{ \tilde{v} \cdot r(\tau_2) \} )
\Delta_{ij} (\sigma_1,\tau_1,\tau_2) 
\Delta_{jk} (\sigma_2,\tau_1,\tau_2) 
\Delta_{ki} (\sigma_3,\tau_1,\tau_2) 
\left. \right],
\nonumber \\ \label{Pform}
\end{eqnarray}
where $P$'s are polynomials with respect to their 
arguments and $\{ \sigma \},\{ \tilde{v} \},
\{ \tilde{v} \cdot r(\tau) \}$ denote the dependence
collectively
\begin{eqnarray}
\{ \sigma \} 
&=& \left\{ \sigma_a ~|~ a \in \{1,2,3 \} \right\},
\nonumber \\
\{ \tilde{v} \}
&=& \left\{ \tilde{v}_a ~|~ a \in \{ij,jk,ki \} \right\},
\nonumber \\
\{ \tilde{v} \cdot r(\tau) \}
&=& \left\{ \tilde{v}_a \cdot r_b(\tau) ~|~ a,b \in \{ij,jk,ki \} \right\}.
\end{eqnarray}
The $P_q$ term contains the whole of the 
diagrams (`$\infty$'-type diagrams) with 4-point vertices. 
The appearance of three propagators in it is due to 
the following transformation:
\begin{eqnarray}
&& \int d \tau d \sigma_2 d\sigma_3
\sum_{i,j,k} 
\Delta_{jk} (\sigma_2,\tau,\tau) 
\Delta_{ki} (\sigma_3,\tau,\tau) 
\nonumber \\
&=& \int d \tau_1 d \tau_2 d\sigma_2 d\sigma_3
\sum_{i,j,k} 
\delta( \tau_1 - \tau_2 )
\Delta_{jk} (\sigma_2,\tau_1,\tau_2) 
\Delta_{ki} (\sigma_3,\tau_1,\tau_2) 
\nonumber \\
&=& \int d \tau_1 d \tau_2 d\sigma_1 d\sigma_2 d\sigma_3
\sum_{i,j,k} 
\{ -\partial_{\sigma_1} \Delta_{ij} (\sigma_1,\tau_1,\tau_2) \}
\Delta_{jk} (\sigma_2,\tau_1,\tau_2) 
\Delta_{ki} (\sigma_3,\tau_1,\tau_2).
\end{eqnarray}
The other terms represent the contributions 
of the diagrams (`$\phi$'-type diagrams) with two 3-point vertices. 
The $P_1$ term contains terms with 
two derivatives operating proper-time propagators,
and the $P_2$ term contains terms with one derivative.
The $P_3$ and $P_4$ terms contain terms 
which do not involve derivatives.
Note that we explicitly separated the terms containing 
$r_{ij}(\tau_1) \cdot r_{ij}(\tau_2)$ as the $P_3$ term from 
the rest of the contributions 
which we called the $P_4$ term.
In deriving the above form, we repeatedly performed 
partial integrations with respect to $\tau_1$ or $\tau_2$,
using the relation
\begin{eqnarray}
r_{ij}^n(\tau) + r_{jk}^n(\tau) + r_{ki}^n(\tau) = 0,
\end{eqnarray}
and also redefining the integration variables,
$\sigma_1,\sigma_2,\sigma_3,\tau_1$ and $\tau_2$,
and indices $i,j$ and $k$ such as
\begin{eqnarray}
i \to j, \quad j \to k, \quad k \to i, \quad
\sigma_1 \to \sigma_2, \quad  \sigma_2 \to \sigma_3, \quad \sigma_3 \to \sigma_1,
\end{eqnarray}
or
\begin{eqnarray}
i \to j, \quad j \to i, \quad
\sigma_2 \to \sigma_3, \quad \sigma_3 \to \sigma_2.
\end{eqnarray}
We will present the explicit forms of
$P_q$, $P_1$, $P_2$, $P_3$ and $P_4$
and some examples of the 
evaluations of contributions from
individual terms in (\ref{Gamma2})
in Appendix A and B, respectively. 

Let us first 
focus on how $\tilde{\Gamma}_{(2)}$ depends on $x_i^n$.
There are two types of dependence:
\begin{enumerate}
\item $ \{ x^2 \}
\equiv \left\{ x_a^2 ~|~ a \in \{ij,jk,ki \} \right\} $,
\item $ \{ x \cdot \tilde{v} \}
\equiv \left\{ x_a \cdot \tilde{v}_b ~|~ a,b \in \{ij,jk,ki \} \right\} $.
\end{enumerate}
The dependence of the type $\{ x^2 \}$ exists only in
 the form 
$ r_{ij}(\tau_1) \cdot r_{ij}(\tau_2)$ except 
inside the proper-time propagators.
The derivatives with respect to $\tau_1$ or $\tau_2$ acting on
proper-time propagators do not induce $\{ x^2 \}$ dependence
as can be seen from the form of 
the proper-time propagator (\ref{Delta}).
The terms with lower powers of $\{ \tilde{v} \}$ must cancel 
in order to agree with the result from the supergravity 
which shows that the 3-body Lagrangian starts from 
the order $O(\tilde{v}^6)$. 
However, it may seem that the cancellations 
cannot be expected {\em before}
the proper-time integrations because of the 
dependence of the type $\{ x^2 \}$ in 
$ r_{ij}(\tau_1) \cdot r_{ij}(\tau_2)$.
In fact, however, 
using the equations 
\begin{eqnarray}
\left[ \partial_{\sigma_1} - \partial_{\tau_1}^2 + r_{ij}(\tau_1)^2 \right]
\Delta_{ij} (\sigma_1,\tau_1,\tau_2) = 0,
\nonumber \\
\left[ \partial_{\sigma_1} - \partial_{\tau_2}^2 + r_{ij}(\tau_2)^2 \right]
\Delta_{ij} (\sigma_1,\tau_1,\tau_2) = 0,
\end{eqnarray}
we can eliminate the factor   
$ r_{ij}(\tau_1) \cdot r_{ij}(\tau_2) $
as 
\begin{eqnarray}
&& ( r_{ij}(\tau_1) \cdot r_{ij}(\tau_2) ) \Delta_{ij} (\sigma_1,\tau_1,\tau_2)
\nonumber \\
&& = \left[
r_{ij}(\tau_1) \cdot r_{ij}(\tau_2)
- \partial_{\sigma_1} 
+ \frac{1}{2} ( \partial_{\tau_1}^2 + \partial_{\tau_2}^2 )
- \frac{1}{2} ( r_{ij}(\tau_1)^2 + r_{ij}(\tau_2)^2 )
\right]
\Delta_{ij} (\sigma_1,\tau_1,\tau_2) 
\nonumber \\
&& = - \partial_{\sigma_1} \Delta_{ij} (\sigma_1,\tau_1,\tau_2)
-\frac{1}{2} ( r_{ij}(\tau_1) - r_{ij}(\tau_2) )^2
\Delta_{ij} (\sigma_1,\tau_1,\tau_2)
\nonumber \\
&& \qquad
+ \frac{1}{2} ( \partial_{\tau_1}^2 + \partial_{\tau_2}^2 )
\Delta_{ij} (\sigma_1,\tau_1,\tau_2).
\end{eqnarray}
 Let us now divide $\tilde{\Gamma}_{(2)}$ into two parts
$\tilde{\Gamma}_{V}$ and $\tilde{\Gamma}_{Y}$
which are expected to correspond to the 
two contributions of the effective Lagrangian 
$L_V$ and $ L_Y$ of the previous section, respectively. 
\begin{eqnarray}
\tilde{\Gamma}_{(2)} = \tilde{\Gamma}_{V} + \tilde{\Gamma}_{Y},
\end{eqnarray}
\begin{eqnarray}
&& -\tilde{\Gamma}_{V} 
\nonumber \\
&& = \frac{g^2}{\kappa} \int d \tau_1 d \tau_2 d\sigma_1 d\sigma_2 d\sigma_3
\sum_{i,j,k} \left[
-\partial_{\sigma_1} \{ ( P_q + P_3 ) \Delta_{ij} (\sigma_1,\tau_1,\tau_2)
\Delta_{jk} (\sigma_2,\tau_1,\tau_2) 
\Delta_{ki} (\sigma_3,\tau_1,\tau_2) \}
\right],
\nonumber \\
\end{eqnarray}
and
\begin{eqnarray}
-\tilde{\Gamma}_{Y} =
\frac{g^2}{\kappa} \int d \tau_1 d \tau_2 d\sigma_1 d\sigma_2 d\sigma_3
\sum_{i,j,k} \left[ \right.
P_1 \{ \partial_{\tau_1} \partial_{\tau_2}
\Delta_{ij} (\sigma_1,\tau_1,\tau_2) \}
\Delta_{jk} (\sigma_2,\tau_1,\tau_2) 
\Delta_{ki} (\sigma_3,\tau_1,\tau_2) 
\nonumber \\
+ P_2 \{ \partial_{\tau_1}
\Delta_{ij} (\sigma_1,\tau_1,\tau_2) \}
\Delta_{jk} (\sigma_2,\tau_1,\tau_2) 
\Delta_{ki} (\sigma_3,\tau_1,\tau_2) 
\nonumber \\
-\frac{1}{2}P_3   ( r_{ij}(\tau_1) - r_{ij}(\tau_2))^2
\Delta_{ij} (\sigma_1,\tau_1,\tau_2)
\Delta_{jk} (\sigma_2,\tau_1,\tau_2) 
\Delta_{ki} (\sigma_3,\tau_1,\tau_2) 
\nonumber \\
+\frac{1}{2}P_3   \{ ( \partial_{\tau_1}^2 + \partial_{\tau_2}^2 )
\Delta_{ij} (\sigma_1,\tau_1,\tau_2) \}
\Delta_{jk} (\sigma_2,\tau_1,\tau_2) 
\Delta_{ki} (\sigma_3,\tau_1,\tau_2) 
\nonumber \\
+ ( P_4 + \partial_{\sigma_1} P_3 ) 
\Delta_{ij} (\sigma_1,\tau_1,\tau_2) 
\Delta_{jk} (\sigma_2,\tau_1,\tau_2) 
\Delta_{ki} (\sigma_3,\tau_1,\tau_2) 
\left. \right].
\nonumber \\
\end{eqnarray}
Since the whole dependence on $\{ x^2 \}$ 
in $\tilde{\Gamma}_{V}$ and $\tilde{\Gamma}_{Y}$
is now contained only in the proper-time propagators, 
 we can expect that the above cancellation could occur before 
the proper-time integrations, that is,
in the prefactors in front of proper-time propagators.  
This is indeed the case. 

\subsubsection{The calculation of $\tilde{\Gamma}_{V}$}
Let us begin with $\tilde{\Gamma}_{V}$.
It is easily calculated as follows:
\begin{eqnarray}
\tilde{\Gamma}_{V} &=& 
- \frac{g^2}{\kappa} \sum_{i,j,k}
\int_{-\infty}^{\infty} d\tau
\int_0^\infty d\sigma_2 \int_0^\infty d\sigma_3 ~
128 \sinh^3 \frac{\sigma_2 \tilde{v}_{jk}}2
\sinh^3 \frac{\sigma_3 \tilde{v}_{ki}}2 
\nonumber \\
&& \quad \times \left( 
\frac{ 2 \tilde{v}_{jk} \cdot \tilde{v}_{ki} }
{\tilde{v}_{jk} \tilde{v}_{ki}}
\cosh \frac{\sigma_2 \tilde{v}_{jk}}2
\cosh \frac{\sigma_3 \tilde{v}_{ki}}2 
- \sinh \frac{\sigma_2 \tilde{v}_{jk}}2
\sinh \frac{\sigma_3 \tilde{v}_{ki}}2 
\right) 
\nonumber \\
&& \quad \times
\Delta_{jk} (\sigma_2,\tau,\tau)
\Delta_{ki} (\sigma_3,\tau,\tau) 
\nonumber \\
&=& 
- \frac{g^2}{\kappa} \sum_{i,j,k}
\int_0^\infty d\sigma_2 \int_0^\infty d\sigma_3 ~
128 \sinh^3 \frac{\sigma_2 \tilde{v}_{jk}}2
\sinh^3 \frac{\sigma_3 \tilde{v}_{ki}}2 
\nonumber \\
&& \quad \times \left( 
\frac{ 2 \tilde{v}_{jk} \cdot \tilde{v}_{ki} }
{\tilde{v}_{jk} \tilde{v}_{ki}}
\cosh \frac{\sigma_2 \tilde{v}_{jk}}2
\cosh \frac{\sigma_3 \tilde{v}_{ki}}2 
- \sinh \frac{\sigma_2 \tilde{v}_{jk}}2
\sinh \frac{\sigma_3 \tilde{v}_{ki}}2 
\right) 
\nonumber \\
&& \quad \times
\sqrt{\frac{\tilde{v}_{jk}}{2 \pi \sinh(2 \sigma_2 \tilde{v}_{jk})}}
\sqrt{\frac{\tilde{v}_{ki}}{2 \pi \sinh(2 \sigma_3 \tilde{v}_{ki})}}
\sqrt{\frac{\pi}{
\tilde{v}_{jk} \tanh (\sigma_2 \tilde{v}_{jk})
+ \tilde{v}_{ki} \tanh (\sigma_3 \tilde{v}_{ki}) }} 
\nonumber \\
&& \quad \times \exp \left[
- \sigma_2 \left(
x_{jk}^2 - \frac{(x_{jk} \cdot \tilde{v}_{jk})^2}{\tilde{v}_{jk}^2}
\right) 
- \sigma_3 \left(
x_{ki}^2 - \frac{(x_{ki} \cdot \tilde{v}_{ki})^2}{\tilde{v}_{ki}^2}
\right) \right. 
\nonumber \\
&& \left. \qquad
- \frac{\tilde{v}_{jk} \, \tilde{v}_{ki}
\tanh (\sigma_2 \tilde{v}_{jk}) \tanh (\sigma_3 \tilde{v}_{ki})}
{\tilde{v}_{jk} \tanh (\sigma_2 \tilde{v}_{jk})
+ \tilde{v}_{ki} \tanh (\sigma_3 \tilde{v}_{ki})}
\left(
\frac{x_{jk} \cdot \tilde{v}_{jk}}{\tilde{v}_{jk}^2}
- \frac{x_{ki} \cdot \tilde{v}_{ki}}{\tilde{v}_{ki}^2}
\right)^2
\right].
\label{GammaV}
\end{eqnarray}
There indeed happened nontrivial cancellations in
the prefactor in front of the proper-time propagators
and it is of order $O( \tilde{v}^6)$.
The expression for $\tilde{\Gamma}_{V}$ (\ref{GammaV})
is exact and
completely general just as 
that for $\tilde{\Gamma}_{(1)}$ (\ref{Gamma1}). 
We want to emphasize that this general 
expression already shows the characteristic kinematical 
structure which is consistent with the result $L_V$ of 
supergravity, and that there 
arises no infrared divergence in the 
limit of coincident D-particles. 

The leading contribution of $\tilde{\Gamma}_{V}$ 
takes the form 
\begin{eqnarray}
\tilde{\Gamma}_{V ~{\rm leading}}
&=& - \frac{g^2}{\kappa} \sum_{i,j,k}
\int_{-\infty}^{\infty} d\tau
\int_0^\infty d\sigma_2 \int_0^\infty d\sigma_3 ~
4 \sigma_2^3 \sigma_3^3 \tilde{v}_{jk}^2 \tilde{v}_{ki}^2
( \tilde{v}_{jk} \cdot \tilde{v}_{ki} ) 
\nonumber \\
&& \qquad \times \frac{1}{4 \pi \sqrt{\sigma_2 \sigma_3}}
\exp \left[
-\sigma_2 ( \tilde{v}_{jk} \tau + x_{jk} )^2
-\sigma_3 ( \tilde{v}_{ki} \tau + x_{ki} )^2
\right] 
\nonumber \\
&=& - \frac{g^2}{\kappa} \sum_{i,j,k} \frac{225}{64}
\int_{-\infty}^{\infty} d\tau
\frac{\tilde{v}_{jk}^2 \tilde{v}_{ki}^2 
( \tilde{v}_{jk} \cdot \tilde{v}_{ki} )}
{[(\tilde{v}_{jk} \tau + x_{jk})^2]^{7/2}
 [(\tilde{v}_{ki} \tau + x_{ki})^2]^{7/2}},
\end{eqnarray}
where we used the proper-time representation (\ref{r7}).

\subsubsection{The calculation of $\tilde{\Gamma}_{Y}$}
We now proceed to $\tilde{\Gamma}_{Y}$.
We can eliminate the time derivatives acting on 
$\Delta_{ij}(\sigma_1,\tau_1,\tau_2)$
using the formulas given in Appendix C. 
The result is then 
\begin{eqnarray}
-\tilde{\Gamma}_{Y} =
\frac{g^2}{\kappa} \int d \tau_1 d \tau_2 d\sigma_1 d\sigma_2 d\sigma_3
\sum_{i,j,k} 
P_Y ( \{ s \} ,\{ \tilde{v} \} ,\{ x \cdot \tilde{v} \},
\tau_1,\tau_2)
\nonumber \\
\times
\Delta_{ij} (\sigma_1,\tau_1,\tau_2) 
\Delta_{jk} (\sigma_2,\tau_1,\tau_2) 
\Delta_{ki} (\sigma_3,\tau_1,\tau_2), 
\end{eqnarray}
where
\begin{eqnarray}
&& P_Y ( \{ s \} ,\{ \tilde{v} \} ,\{ x \cdot \tilde{v} \},
\tau_1,\tau_2) 
\nonumber \\
&& \qquad = P_1 \left[
-\frac{\tilde{v}_{ij}}{2} \frac{V_{ij}(\sigma_1)}{C_{ij}(\sigma_1)} 
+\frac{\tilde{v}_{ij}}{2} \frac{C_{ij}(\sigma_1)}{V_{ij}(\sigma_1)}
+ \left( \frac{V_{ij}(\sigma_1)}{C_{ij}(\sigma_1)} \cdot \frac{r_{ij}(\tau_1)+r_{ij}(\tau_2)}{2} \right)^2 \right.
\nonumber \\
&& \qquad \qquad \quad \left.
- \left( \frac{C_{ij}(\sigma_1)}{V_{ij}(\sigma_1)} \frac{r_{ij}(\tau_1)-r_{ij}(\tau_2)}{2} \right)^2
\right] 
\nonumber \\
&& \qquad \quad + P_2 \left[
- \frac{V_{ij}(\sigma_1)}{C_{ij}(\sigma_1)} \cdot \frac{r_{ij}(\tau_1)+r_{ij}(\tau_2)}{2}
- \frac{C_{ij}(\sigma_1)}{V_{ij}(\sigma_1)} \frac{|r_{ij}(\tau_1)-r_{ij}(\tau_2)|}{2}
\right] 
\nonumber \\
&& \qquad \quad + P_3 \left[
-2 \left( \frac{r_{ij}(\tau_1)-r_{ij}(\tau_2)}{2} \right)^2
-\frac{\tilde{v}_{ij}}{2} \frac{V_{ij}(\sigma_1)}{C_{ij}(\sigma_1)} 
-\frac{\tilde{v}_{ij}}{2} \frac{C_{ij}(\sigma_1)}{V_{ij}(\sigma_1)} \right.
\nonumber \\
&& \qquad \qquad \quad \left.
+ \left( \frac{V_{ij}(\sigma_1)}{C_{ij}(\sigma_1)} \cdot \frac{r_{ij}(\tau_1)+r_{ij}(\tau_2)}{2} \right)^2
+ \left( \frac{C_{ij}(\sigma_1)}{V_{ij}(\sigma_1)} \frac{r_{ij}(\tau_1)-r_{ij}(\tau_2)}{2} \right)^2
\right] 
\nonumber \\
&& \qquad \quad + P_4 + \partial_{\sigma_1} P_3.
\label{Py}
\end{eqnarray}
We have made no approximations for $\tilde{\Gamma}_{Y}$ so far
and we can carry out the integrations
with respect to $\tau_1$ and $\tau_2$ as well.
Thus, we can in principle evaluate the leading contribution
with respect to $\{ \tilde{v} \}$ simply 
by expanding it to Taylor series.
However, the expression is too complicated to do that
in a completely general way.
So we content ourselves to restrict  to the cases
where $\{ x \cdot \tilde{v} \}$ vanish.  Since the relative 
velocities $\tilde{v}_{ij}$ are independent of  the relative 
coordinates $x_{ij}$ at initial time $\tau=0$ ,
 we can arrange, for arbitrary  given configurations of 
relative velocities, 
the trajectories of D-particles by making  parallel transport  of  
the positions of each D-particle  at time $\tau=0$ 
 appropriately  such that the conditions  
are satisfied for  three D-particles   
 which participate in the 3-body interaction 
in 9 transverse dimensions.     
Then the integrations
with respect to $\tau_1$ and $\tau_2$ are considerably simplified.
Using the formulas in Appendix C, 
$\tilde{\Gamma}_{Y}$ can be rearranged into the form 
\begin{eqnarray}
-\tilde{\Gamma}_{Y} =
\frac{g^2}{\kappa} \int d \tau_1 d \tau_2 d\sigma_1 d\sigma_2 d\sigma_3
\sum_{i,j,k} 
\tilde{P}_Y(\sigma_1,\sigma_2,\sigma_3,\tilde{v}_{ij},\tilde{v}_{jk},\tilde{v}_{ki})
\nonumber \\ \times
\Delta_{ij} (\sigma_1,\tau_1,\tau_2) 
\Delta_{jk} (\sigma_2,\tau_1,\tau_2) 
\Delta_{ki} (\sigma_3,\tau_1,\tau_2),
\end{eqnarray}
where
\begin{eqnarray}
&& \tilde{P}_Y(\sigma_1,\sigma_2,\sigma_3,\tilde{v}_{ij},\tilde{v}_{jk},\tilde{v}_{ki}) 
= P_1 \left[
-\frac{\tilde{v}_{ij}}{2} \frac{V_{ij}(\sigma_1)}{C_{ij}(\sigma_1)} 
+\frac{\tilde{v}_{ij}}{2} \frac{C_{ij}(\sigma_1)}{V_{ij}(\sigma_1)}
\right]
\nonumber \\
&& \qquad + P_3 \left[
-\frac{\tilde{v}_{ij}}{2} \frac{V_{ij}(\sigma_1)}{C_{ij}(\sigma_1)} 
-\frac{\tilde{v}_{ij}}{2} \frac{C_{ij}(\sigma_1)}{V_{ij}(\sigma_1)}
\right]
+ \partial_{\sigma_1} P_3 
\nonumber \\
&& \qquad + \frac{1}{2} \left( 
\frac{\tilde{v}_{ij} V_{ij}(\sigma_1)}{C_{ij}(\sigma_1)} + \frac{\tilde{v}_{jk} V_{jk}(\sigma_2)}{C_{jk}(\sigma_2)} + \frac{\tilde{v}_{ki} V_{ki}(\sigma_3)}{C_{ki}(\sigma_3)}
\right)^{-1}
\nonumber \\
&& \qquad \qquad \left[
P_1 \left( \frac{\tilde{v}_{ij} V_{ij}(\sigma_1)}{C_{ij}(\sigma_1)} \right)^2
- \frac{P_2}{\tau_2} \frac{\tilde{v}_{ij} V_{ij}(\sigma_1)}{C_{ij}(\sigma_1)}
+ P_3 
\left( \frac{\tilde{v}_{ij} V_{ij}(\sigma_1)}{C_{ij}(\sigma_1)} \right)^2
+ \frac{P_4}{\tau_1 \tau_2}
\right] 
\nonumber \\
&& \qquad + \frac{1}{2} \left( 
\frac{\tilde{v}_{ij} C_{ij}(\sigma_1)}{V_{ij}(\sigma_1)} + \frac{\tilde{v}_{jk} C_{jk}(\sigma_2)}{V_{jk}(\sigma_2)} + \frac{\tilde{v}_{ki} C_{ki}(\sigma_3)}{V_{ki}(\sigma_3)}
\right)^{-1}
\nonumber \\
&& \qquad \qquad \left[
- P_1 
\left( \frac{\tilde{v}_{ij} C_{ij}(\sigma_1)}{V_{ij}(\sigma_1)} \right)^2
+ \frac{P_2}{\tau_2} \frac{\tilde{v}_{ij} C_{ij}(\sigma_1)}{V_{ij}(\sigma_1)}
- 2 \tilde{v}_{ij}^2P_3  
+ P_3 \left( \frac{\tilde{v}_{ij} C_{ij}(\sigma_1)}{V_{ij}(\sigma_1)} \right)^2
- \frac{P_4}{\tau_1 \tau_2}
\right].
\nonumber \\
\end{eqnarray}
We expand $\tilde{P}_Y$
with respect to $\tilde{v}_{ij}$, $\tilde{v}_{jk}$ and $\tilde{v}_{ki}$ 
after symmetrization. 
The result is
\footnote{We performed this computation 
with the help of the symbolic computation program 
Maple V Release 4. }
\begin{eqnarray}
&& \tilde{P}_Y(\sigma_1,\sigma_2,\sigma_3,\tilde{v}_{ij},\tilde{v}_{jk},\tilde{v}_{ki})_{sym}
\nonumber \\
&& =\frac{1}{6} [
\tilde{P}_Y(\sigma_1,\sigma_2,\sigma_3,\tilde{v}_{ij},\tilde{v}_{jk},\tilde{v}_{ki})
+ \tilde{P}_Y(\sigma_1,\sigma_3,\sigma_2,\tilde{v}_{ij},\tilde{v}_{ki},\tilde{v}_{jk})
+ \tilde{P}_Y(\sigma_2,\sigma_1,\sigma_3,\tilde{v}_{jk},\tilde{v}_{ij},\tilde{v}_{ki}) 
\nonumber \\
&& \qquad 
+ \tilde{P}_Y(\sigma_2,\sigma_3,\sigma_1,\tilde{v}_{jk},\tilde{v}_{ki},\tilde{v}_{ij})
+ \tilde{P}_Y(\sigma_3,\sigma_1,\sigma_2,\tilde{v}_{ki},\tilde{v}_{ij},\tilde{v}_{jk})
+ \tilde{P}_Y(\sigma_3,\sigma_2,\sigma_1,\tilde{v}_{ki},\tilde{v}_{jk},\tilde{v}_{ij})
] 
\nonumber \\
&& = -\frac{1}{6}
( \tilde{v}_{ij} + \tilde{v}_{jk} + \tilde{v}_{ki})
(-\tilde{v}_{ij} + \tilde{v}_{jk} + \tilde{v}_{ki})
( \tilde{v}_{ij} - \tilde{v}_{jk} + \tilde{v}_{ki})
( \tilde{v}_{ij} + \tilde{v}_{jk} - \tilde{v}_{ki}) 
\nonumber \\
&&
\qquad \times ( \sigma_1 \sigma_2 + \sigma_2 \sigma_3 + \sigma_3 \sigma_1 )^2
( \sigma_1 \tilde{v}_{ij}^2 + \sigma_2 \tilde{v}_{jk}^2 + \sigma_3 \tilde{v}_{ki}^2 )
+ O(\tilde{v}^8).
\end{eqnarray}
We again encountered miraculous cancellations here
just as in $\tilde{\Gamma}_{V}$.
Then, after integrating over the 
times $\tau_1$ and $\tau_2$, 
the leading contribution to $\tilde{\Gamma}_{Y}$ is given by 
\begin{eqnarray}
\tilde{\Gamma}_{Y ~{\rm leading}} &=&
\frac{g^2}{\kappa} \sum_{i,j,k}
\int_{0}^{\infty} d \sigma_1
\int_{0}^{\infty} d \sigma_2
\int_{0}^{\infty} d \sigma_3
~\frac{1}{24 \sqrt{\pi}}
( \tilde{v}_{ij} + \tilde{v}_{jk} + \tilde{v}_{ki} ) 
\nonumber \\
&& \qquad \times
(-\tilde{v}_{ij} + \tilde{v}_{jk} + \tilde{v}_{ki} )
( \tilde{v}_{ij} - \tilde{v}_{jk} + \tilde{v}_{ki} )
( \tilde{v}_{ij} + \tilde{v}_{jk} - \tilde{v}_{ki} ) 
\nonumber \\
&& \qquad \times
( \sigma_1 \sigma_2 + \sigma_2 \sigma_3 + \sigma_3 \sigma_1 )^{3/2}
\sqrt{\sigma_1 \tilde{v}_{ij}^2 + \sigma_2 \tilde{v}_{jk}^2 + \sigma_3 \tilde{v}_{ki}^2} 
\nonumber \\
&& \qquad \times
\exp (-\sigma_1 x_{ij}^2 -\sigma_2 x_{jk}^2 -\sigma_3 x_{ki}^2 ).
\end{eqnarray}
This formula vanishes whenever any one of the relative velocities 
vanishes, since the product 
\[{1\over 16} \, 
( \tilde{v}_{ij} + \tilde{v}_{jk} + \tilde{v}_{ki} ) 
(-\tilde{v}_{ij} + \tilde{v}_{jk} + \tilde{v}_{ki} )
( \tilde{v}_{ij} - \tilde{v}_{jk} + \tilde{v}_{ki} )
( \tilde{v}_{ij} + \tilde{v}_{jk} - \tilde{v}_{ki} ) 
\]
is just the square of the area, known as Heron's formula, 
of the triangle 
formed by three relative velocity vectors 
$\tilde{v}_{ij} , \tilde{v}_{jk} , \tilde{v}_{ki}$. 
This is a nontrivial check of our result 
in conformity with the BPS property of D-particles 
and consistent with the structure of the 
3-body Lagrangian of supergravity. 


\section{Comparison between supergravity and Matrix theory}
\setcounter{equation}{0}
We are now ready to compare the results of Matrix theory 
with the prediction of supergravity. 
To compare the effective action 
of the Matrix-theory calculations
with the result from supergravity,
we have first to take into account 
the changes of conventions in both 
cases, most of which are already explained 
in the beginning of section 3 (see below (\ref{Bdef})).  
Let us summarize them again here. 

First, while we computed the effective action 
with completely general diagonal backgrounds,
we are actually interested in backgrounds consisting  of
several blocks proportional to unit matrices of the order $N_a$
($a=1,2,\ldots$, $N=\sum_a N_a$).
The summations over U($N$) indices automatically
take care of this effect and just reduce to the summation 
over the block elements.
Secondly, since the supergravity calculations are
carried out in Minkowski space-time, we have to transform
the Euclidean effective action to Minkowski signature.
Thirdly, we have to rescale the coordinates and velocities: 
 \begin{eqnarray}
&& \sum_{i} \to \sum_{a} N_a, \quad
\tilde{\Gamma} \to -i \Gamma, \quad
\tau \to it, \quad
\kappa \to \frac{1}{R}, \quad
g \to M^3 R, \quad
\nonumber \\
&& x \to gx = M^3 R x, \quad
\tilde{v} \to g(-iv) = M^3 R (-iv), \quad
\sigma \to \frac{\sigma}{g^2} = \frac{\sigma}{M^6 R^2}.
\end{eqnarray}

With these replacements,  the Matrix-theory results read 
\begin{eqnarray}
\Gamma_{(0)} 
&=& \int_{-\infty}^\infty dt
\sum_a \frac{1}{2} \frac{N_a}{R} v_a^2 
\\
\Gamma_{(1) ~{\rm leading}}
&=& \int_{-\infty}^\infty dt \sum_{a<b}
\frac{15}{16} \frac{N_a N_b}{R^3 M^9}
\frac{v_{ab}^4}
{[(v_{ab} t + x_{ab})^2]^{7/2}},
\label{phase1} \\
\Gamma_{V ~{\rm leading}}
&=& - \int_{-\infty}^{\infty} dt
\sum_{a,b,c} \frac{225}{64}
\frac{N_a N_b N_c}{R^5 M^{18}}
\frac{v_{bc}^2 v_{ca}^2 
( v_{bc} \cdot v_{ca} )}
{[(v_{bc} t + x_{bc})^2]^{7/2}
 [(v_{ca} t + x_{ca})^2]^{7/2}},
\label{phaseV} \\
\Gamma_{Y ~{\rm leading}} 
&=&
\int_{0}^{\infty} d \sigma_1
\int_{0}^{\infty} d \sigma_2
\int_{0}^{\infty} d \sigma_3
~ \sum_{a,b,c}
\frac{1}{24 \sqrt{\pi}}
\frac{N_a N_b N_c}{R^5 M^{18}}
( v_{ab} + v_{bc} + v_{ca} ) 
\nonumber \\
&& \qquad \times
(-v_{ab} + v_{bc} + v_{ca} )
( v_{ab} - v_{bc} + v_{ca} )
( v_{ab} + v_{bc} - v_{ca} )
\nonumber \\
&& \qquad \times
( \sigma_1 \sigma_2 + \sigma_2 \sigma_3 + \sigma_3 \sigma_1 )^{3/2}
\sqrt{\sigma_1 v_{ab}^2 + \sigma_2 v_{bc}^2 + \sigma_3 v_{ca}^2} 
\nonumber \\
&& \qquad \times
\exp (-\sigma_1 x_{ab}^2 -\sigma_2 x_{bc}^2 -\sigma_3 x_{ca}^2 ).
\end{eqnarray}
Recall that the results $\Gamma_{(0)}$, 
$\Gamma_{(1) ~{\rm leading}}$ and 
$\Gamma_{V ~{\rm leading}}$ are for completely general situations, 
while $\Gamma_{Y ~{\rm leading}}$ is  for the 
case where $\{ x \cdot v\}$ vanish.  Note, however, as already 
emphasized before, that 
the latter condition can be satisfied for arbitrary 
configurations of relative velocities of three D-particles,  
provided that  only the 
relative coordinates at initial time are chosen appropriately. 

We turn to computations of the phase shift 
in the eikonal approximation 
using the effective Lagrangian obtained in supergravity .
As for the 2-body Lagrangian $L_2$ (\ref{eq253}) 
and the $L_V$ part (\ref{eq256}) of the 3-body Lagrangian,
it is already evident that they reproduce
exactly the same phase shifts as those of Matrix theory
(\ref{phase1}) and (\ref{phaseV}), 
since the integrands in these expressions 
are identical with the corresponding effective Lagrangians 
of supergravity.   

For the $L_Y$ part, the result in Matrix theory is 
given in a form where the integration over time 
has been performed.  Thus the relation with 
supergravity is not obvious. 
Let us therefore explicitly compute the eikonal phase shift from
the $L_Y$ part of the 3-body Lagrangian. 
We compute it with the same restriction $\{ x \cdot v\}=0$ 
as we have made for $\Gamma_{Y}$ in Matrix theory, 
assuming the straight-line trajectories $x_a(\tau) 
= x_a + v_a \tau$. 
Then by a tedious but straightforward calculation,  including
symmetrization with respect to the block 
indices $a$,$b$ and $c$,  we can arrange 
the time integral of the Lagrangian $L_Y$ to the following form
\begin{eqnarray}
&&-\int_{-\infty}^{\infty} d \tau 
\sum_{a,b,c} {(15)^3N_a N_b N_c
\over 96(2\pi)^4} \Bigl[
-v_{bc}^2 v_{ca}^2(v_{cb}\cdot \nabla_c)
(v_{ca}\cdot \nabla_c)   \nonumber\\
&&+ {1\over 2}v_{ca}^4(v_{cb}\cdot \nabla_c)^2  
+{1\over 2}v_{bc}^4 (v_{ca}\cdot \nabla_c)^2
 -{1\over 2}v_{ba}^2
v_{ac}^2(v_{cb}\cdot \nabla_c)(v_{bc}\cdot \nabla_b)
 \nonumber\\
&&+{1\over 4}v_{bc}^4 
( v_{ba}\cdot \nabla_b)(v_{ca}\cdot \nabla_c)
\Bigr]\Delta(a,b,c) 
\nonumber \\
 &=& \int_{-\infty}^{\infty} d \tau \sum_{a,b,c}  
{N_a N_b N_c
\over 24\pi}\int_{0}^{\infty} d^3 \sigma \, 
( \sigma_1 \sigma_2 + \sigma_2 \sigma_3 + \sigma_3 \sigma_1 )^{3/2} 
\nonumber \\
&& \qquad \times
( v_{ab} + v_{bc} + v_{ca} )  
(-v_{ab} + v_{bc} + v_{ca} )
( v_{ab} - v_{bc} + v_{ca} )
( v_{ab} + v_{bc} - v_{ca} )
\nonumber \\
&& \qquad \times
(\sigma_1 v_{ab}^2 + \sigma_2 v_{bc}^2 + \sigma_3 v_{ca}^2) 
\exp (-\sigma_1 x_{ab}^2 -\sigma_2 x_{bc}^2 -\sigma_3 x_{ca}^2 ).
\label{eq46}
\end{eqnarray}
Performing the time integral explicitly
 gives the following expression
for the phase shift:
\begin{eqnarray}
&& \int_{0}^{\infty} d \sigma_1
\int_{0}^{\infty} d \sigma_2
\int_{0}^{\infty} d \sigma_3
~ \sum_{a,b,c}
\frac{1}{24 \sqrt{\pi}}
\frac{N_a N_b N_c}{R^5 M^{18}}
( v_{ab} + v_{bc} + v_{ca} ) 
\nonumber \\
&& \qquad \times
(-v_{ab} + v_{bc} + v_{ca} )
( v_{ab} - v_{bc} + v_{ca} )
( v_{ab} + v_{bc} - v_{ca} )
\nonumber \\
&& \qquad \times
( \sigma_1 \sigma_2 + \sigma_2 \sigma_3 + \sigma_3 \sigma_1 )^{3/2}
\sqrt{\sigma_1 v_{ab}^2 + \sigma_2 v_{bc}^2 + \sigma_3 v_{ca}^2} 
\nonumber \\
&& \qquad \times
\exp (-\sigma_1 x_{ab}^2 -\sigma_2 x_{bc}^2 -\sigma_3 x_{ca}^2 ),
\end{eqnarray}
which precisely coincides with $\Gamma_{Y~{\rm leading}}$ from Matrix theory 
including the sign and numerical coefficient.  
Since, in this calculation, all the terms in the left hand side in 
(\ref{eq46}) contribute in a nontrivial way,   
we are confident that the agreement will 
extend to the general case where the technical restriction 
$\{x\cdot  v\}=0$ is not assumed. 

It should be emphasized that in both of our supergravity and 
Matrix theory calculations, the recoil effect is neglected completely. 
In order to take into account the 
recoil effect, we have to allow non-local (in time) contributions  
to the effective action in supergravity and 
the 1-particle reducible contributions  
in the phase shift of Matrix theory.  We hope to report 
on the recoil effect in a forthcoming work. 

\section{Concluding Remarks}
We have presented a detailed study of the 
3-body interaction of D-particles and established 
its precise equivalence in Matrix theory and 11 dimensional 
supergravity. Our results provide a strong support for the 
idea \cite{susskind} that the Matrix theory is a formulation of M-theory in 
the discrete light-cone quantization.  As emphasized in 
the Introduction, we also view our results as evidence for Lorentz 
invariance of  Matrix theory.  In a sense, what we have done 
in the present paper may be regarded as a Matrix-theory 
analogue to the old works 
\cite{ty}\cite{ss} in which the 
non-linear graviton behavior in string theory was first 
established.  In the case of the usual string theory, 
the formalism is Lorentz invariant from the 
beginning and hence the emergence of 
Einstein gravity is ensured by checking the 
non-linear effect to the first nontrivial 
order,  owing to the general theorem (see, e. g. \cite{wyss}) 
showing the uniqueness of the  Einstein equation.  
Because of this, we could be 
fairly  confident in that the string theory 
describes gravity in a consistent way in the 
long-distance limit. In the case of Matrix theory, 
unfortunately,  Lorentz invariant formulation still 
remains as a big mystery.  
Perhaps one of the most crucial issue 
in the connection of Matrix theory and supergravity 
is to establish 
a more systematic way of making correspondence 
between supergravity and Matrix theory. 
 Technically, for example, it may be 
possible to relate both theories order by order at the 
level of Feynman rules. 
There should be some correspondence principle 
which ensures that the long-distance 
property of Matrix theory is reproduced by 
supergravity.    

Finally, in view of the recent discovery \cite{jty} of the 
conformal symmetry associated with the 
space-time uncertainty relation for D-particles \cite{lty},  
it is interesting to see whether there exists 
a generalization of the conformal symmetry for 
our effective action for many D-particle systems. 
Such a symmetry, if it exists, combined with the 
supersymmetry is expected to determine the 
form of the effective action to all orders in the  
velocity expansion in the classical limit.  
A related problem is the supersymmetric generalization 
of our effective action to include the spin degrees of freedom.  

\vspace{0.5cm}
\noindent
{\it Note added}

After the completion of the present work,
 in the course of preparing the  manuscript, 
two papers \cite{fabb}\cite{taylor} 
discussing the same subject appeared in hep-th archive. 
However, both of them reported only 
the considerations of particular subsets of the 
relevant contributions and hence remain inconclusive 
as to the agreement between supergravity and 
Matrix theory. We believe that the present work 
gives a final resolution on the issue.  

\vspace{1cm}
\noindent
Acknowledgements

We would like to thank M. Ikehara for discussions on two-loop 
computations in matrix models and, especially, to 
Y. Kazama for collaborative discussions on the role of supersymmetry 
in the dynamics of D-particles at an early stage of the 
present work.  
A part of the present work was done during one (T. Y.) of the 
authors was staying at ITP, Santa Barbara. T.Y. would like to 
thank ITP for its hospitality.  
The research at ITP was supported in part by the National 
Science Foundation under Grant No. PHY94-07194.  
The work of T.Y. is supported in part 
by Grant-in-Aid for Scientific  Research (No. 09640337) 
and Grant-in-Aid for International Scientific Research 
(Joint Research, No. 10044061) from the Ministry of  Education, Science and Culture.  The work of Y. O. is supported 
in part by the Japan Society for the Promotion of 
Science under the Predoctoral Research Program (No. 08-4158). 
 

\newpage
\noindent
{\large Appendix}
\renewcommand{\theequation}{\Alph{section}.\arabic{equation}}

\appendix
\section{Explicit forms of $P_q$, $P_1$, $P_2$, $P_3$ and $P_4$}
\setcounter{equation}{0}
We here present the explicit forms of $P_q$, $P_1$, $P_2$, $P_3$ and $P_4$.
\begin{eqnarray}
P_q &=&
-45 -18 V_{jk}(\sigma_2)^2 -18 V_{ki}(\sigma_3)^2
+ 12 C_{jk}(\sigma_2) C_{ki}(\sigma_3) ( V_{jk}(\sigma_2) \cdot V_{ki}(\sigma_3) )
\nonumber \\
&& -6 V_{jk}(\sigma_2)^2 V_{ki}(\sigma_3)^2 + 2 ( V_{jk}(\sigma_2) \cdot V_{ki}(\sigma_3) )^2,
\end{eqnarray}
\begin{eqnarray}
P_1 &=& 
-(1+2 V_{ij}(\sigma_1)^2) 
\{ 4+ V_{jk}(\sigma_2)^2 + V_{ki}(\sigma_3)^2 + 2 ( V_{jk}(\sigma_2) \cdot V_{ki}(\sigma_3) )^2 \} 
\nonumber \\
&& +(1+2 V_{jk}(\sigma_2)^2)
\left\{ \frac{17}{2} + 2 V_{ij}(\sigma_1)^2 + 2 V_{ki}(\sigma_3)^2 + 4 ( V_{ij}(\sigma_1) \cdot V_{ki}(\sigma_3) )^2 
\right\} 
\nonumber \\
&& +(1+2 V_{ki}(\sigma_3)^2) \left\{
\frac{17}{2} + 2 V_{ij}(\sigma_1)^2 + 2 V_{jk}(\sigma_2)^2 + 4 ( V_{ij}(\sigma_1) \cdot V_{jk}(\sigma_2) )^2 
\right\} 
\nonumber \\
&& -10 C_{jk}(\sigma_2) C_{ki}(\sigma_3) \{ V_{jk}(\sigma_2) \cdot V_{ki}(\sigma_3)
+ 2 ( V_{ij}(\sigma_1) \cdot V_{jk}(\sigma_2) ) ( V_{ij}(\sigma_1) \cdot V_{ki}(\sigma_3) ) \} 
\nonumber \\
&& + 2 C_{ij}(\sigma_1) C_{jk}(\sigma_2) \{ V_{ij}(\sigma_1) \cdot V_{jk}(\sigma_2) 
+ 2 ( V_{ij}(\sigma_1) \cdot V_{ki}(\sigma_3) ) ( V_{jk}(\sigma_2) \cdot V_{ki}(\sigma_3) ) \} 
\nonumber \\
&& + 2 C_{ij}(\sigma_1) C_{ki}(\sigma_3) \{ V_{ij}(\sigma_1) \cdot V_{ki}(\sigma_3) 
+ 2 ( V_{ij}(\sigma_1) \cdot V_{jk}(\sigma_2) ) ( V_{jk}(\sigma_2) \cdot V_{ki}(\sigma_3) ) \} 
\nonumber \\
&& + 16 \{
-2 C_{jk}(\sigma_2) C_{ki}(\sigma_3) -2 V_{jk}(\sigma_2) \cdot V_{ki}(\sigma_3) -V_{ij}(\sigma_1)^2 ( V_{jk}(\sigma_2) \cdot V_{ki}(\sigma_3) )
\nonumber \\
&& +( V_{ij}(\sigma_1) \cdot V_{jk}(\sigma_2) ) ( V_{ij}(\sigma_1) \cdot V_{ki}(\sigma_3) ) 
\nonumber \\
&& +2 C_{ij}(\sigma_1) C_{ki}(\sigma_3) +2 V_{ij}(\sigma_1) \cdot V_{ki}(\sigma_3) +V_{jk}(\sigma_2)^2 ( V_{ij}(\sigma_1) \cdot V_{ki}(\sigma_3) )
\nonumber \\
&& -( V_{ij}(\sigma_1) \cdot V_{jk}(\sigma_2) ) ( V_{jk}(\sigma_2) \cdot V_{ki}(\sigma_3) ) 
\nonumber \\
&& +2 C_{ij}(\sigma_1) C_{jk}(\sigma_2) +2 V_{ij}(\sigma_1) \cdot V_{jk}(\sigma_2) +V_{ki}(\sigma_3)^2 ( V_{ij}(\sigma_1) \cdot V_{jk}(\sigma_2) )
\nonumber \\
&& -( V_{ij}(\sigma_1) \cdot V_{ki}(\sigma_3) ) ( V_{jk}(\sigma_2) \cdot V_{ki}(\sigma_3) ) \},
\end{eqnarray}
\begin{eqnarray}
P_2 &=& 
-4 C_{ij}(\sigma_1) ( V_{ij}(\sigma_1) \cdot r_{ij}(\tau_2) ) 
\nonumber \\
&& \qquad \times
\{ 4+ V_{jk}(\sigma_2)^2 + V_{ki}(\sigma_3)^2 + 2 ( V_{jk}(\sigma_2) \cdot V_{ki}(\sigma_3) )^2 
\nonumber \\
&& -2 C_{jk}(\sigma_2) C_{ki}(\sigma_3) ( V_{jk}(\sigma_2) \cdot V_{ki}(\sigma_3) ) \} 
\nonumber \\
&& +4 C_{jk}(\sigma_2) ( V_{jk}(\sigma_2) \cdot r_{ij}(\tau_2) ) 
\nonumber \\
&& \qquad \times \left\{
\frac{17}{2} + 2 V_{ij}(\sigma_1)^2 + 2 V_{ki}(\sigma_3)^2 + 4 ( V_{ij}(\sigma_1) \cdot V_{ki}(\sigma_3) )^2 \right.
\nonumber \\
&& \left. 
-4 C_{ij}(\sigma_1) C_{ki}(\sigma_3) ( V_{ij}(\sigma_1) \cdot V_{ki}(\sigma_3) ) \right\} 
\nonumber \\
&& +4 C_{ki}(\sigma_3) ( V_{ki}(\sigma_3) \cdot r_{ij}(\tau_2) ) 
\nonumber \\
&& \qquad \times \left\{
\frac{17}{2} + 2 V_{ij}(\sigma_1)^2 + 2 V_{jk}(\sigma_2)^2 + 4 ( V_{ij}(\sigma_1) \cdot V_{jk}(\sigma_2) )^2 \right.
\nonumber \\
&& \left.
-4 C_{ij}(\sigma_1) C_{jk}(\sigma_2) ( V_{ij}(\sigma_1) \cdot V_{jk}(\sigma_2) ) \right\} 
\nonumber \\
&& +4 C_{jk}(\sigma_2) V_{jk}(\sigma_2) \cdot ( r_{jk}(\tau_2) - r_{ij}(\tau_2) )
+4 C_{ki}(\sigma_3) V_{ki}(\sigma_3) \cdot ( r_{jk}(\tau_2) - r_{ki}(\tau_2) ) 
\nonumber \\
&& +4 \{ C_{ij}(\sigma_1) -2 C_{jk}(\sigma_2) ( C_{ij}(\sigma_1) C_{jk}(\sigma_2) - V_{ij}(\sigma_1) \cdot V_{jk}(\sigma_2) ) \}
V_{ij}(\sigma_1) \cdot ( r_{jk}(\tau_2) - r_{ij}(\tau_2) ) 
\nonumber \\
&& +4 \{ C_{jk}(\sigma_2) -2 C_{ki}(\sigma_3) ( C_{jk}(\sigma_2) C_{ki}(\sigma_3) - V_{jk}(\sigma_2) \cdot V_{ki}(\sigma_3) ) \}
V_{jk}(\sigma_2) \cdot ( r_{jk}(\tau_2) - r_{ki}(\tau_2) ) 
\nonumber \\
&& +8 \{ C_{jk}(\sigma_2) ( V_{jk}(\sigma_2) \cdot V_{ki}(\sigma_3) ) + C_{ij}(\sigma_1) ( V_{ij}(\sigma_1) \cdot V_{ki}(\sigma_3) )
\nonumber \\
&& -2 C_{jk}(\sigma_2) ( V_{ij}(\sigma_1) \cdot V_{ki}(\sigma_3) ) ( C_{ij}(\sigma_1) C_{jk}(\sigma_2) - V_{ij}(\sigma_1) \cdot V_{jk}(\sigma_2) ) \}
\nonumber \\
&& \qquad \times V_{ki}(\sigma_3) \cdot ( r_{jk}(\tau_2) - r_{ij}(\tau_2) ) 
\nonumber \\
&& +8 \{ C_{ki}(\sigma_3) ( V_{ij}(\sigma_1) \cdot V_{ki}(\sigma_3) ) + C_{jk}(\sigma_2) ( V_{ij}(\sigma_1) \cdot V_{jk}(\sigma_2) )
\nonumber \\
&& -2 C_{ki}(\sigma_3) ( V_{ij}(\sigma_1) \cdot V_{jk}(\sigma_2) ) ( C_{jk}(\sigma_2) C_{ki}(\sigma_3) - V_{jk}(\sigma_2) \cdot V_{ki}(\sigma_3) ) \}
\nonumber \\
&& \qquad \times V_{ij}(\sigma_1) \cdot ( r_{jk}(\tau_2) - r_{ki}(\tau_2) ) 
\nonumber \\
&& +64 \{ -2 C_{ij}(\sigma_1) ( V_{jk}(\sigma_2) \cdot r_{jk}(\tau_2) ) 
- C_{jk}(\sigma_2) ( V_{ij}(\sigma_1) \cdot r_{jk}(\tau_2) )
\nonumber \\
&& - C_{ki}(\sigma_3) ( C_{jk}(\sigma_2) C_{ki}(\sigma_3) - V_{jk}(\sigma_2) \cdot V_{ki}(\sigma_3) ) ( V_{ij}(\sigma_1) \cdot r_{jk}(\tau_2) ) 
\nonumber \\
&& + C_{jk}(\sigma_2) ( V_{ij}(\sigma_1) \cdot V_{ki}(\sigma_3) ) ( V_{ki}(\sigma_3) \cdot r_{jk}(\tau_2) )
\nonumber \\
&& - C_{ki}(\sigma_3) ( V_{ij}(\sigma_1) \cdot V_{jk}(\sigma_2) ) ( V_{ki}(\sigma_3) \cdot r_{jk}(\tau_2) ) \},
\end{eqnarray}
\begin{eqnarray}
P_3 &=& 
7 -2 ( C_{jk}(\sigma_2) C_{ki}(\sigma_3) - V_{jk}(\sigma_2) \cdot V_{ki}(\sigma_3) )^2
\nonumber \\
&& +4 ( C_{ij}(\sigma_1) C_{jk}(\sigma_2) - V_{ij}(\sigma_1) \cdot V_{jk}(\sigma_2) )^2 
+4 ( C_{ij}(\sigma_1) C_{ki}(\sigma_3) - V_{ij}(\sigma_1) \cdot V_{ki}(\sigma_3) )^2 
\nonumber \\
&& + 16 \{
- ( C_{jk}(\sigma_2) C_{ki}(\sigma_3) - V_{jk}(\sigma_2) \cdot V_{ki}(\sigma_3) )
\nonumber \\
&& - ( C_{ij}(\sigma_1) C_{jk}(\sigma_2) - V_{ij}(\sigma_1) \cdot V_{jk}(\sigma_2) ) ( C_{ij}(\sigma_1) C_{ki}(\sigma_3) - V_{ij}(\sigma_1) \cdot V_{ki}(\sigma_3) ) 
\nonumber \\
&& + ( C_{ij}(\sigma_1) C_{jk}(\sigma_2) - V_{ij}(\sigma_1) \cdot V_{jk}(\sigma_2) )
\nonumber \\
&& + ( C_{ij}(\sigma_1) C_{ki}(\sigma_3) - V_{ij}(\sigma_1) \cdot V_{ki}(\sigma_3) ) ( C_{jk}(\sigma_2) C_{ki}(\sigma_3) - V_{jk}(\sigma_2) \cdot V_{ki}(\sigma_3) ) 
\nonumber \\
&& + ( C_{ij}(\sigma_1) C_{ki}(\sigma_3) - V_{ij}(\sigma_1) \cdot V_{ki}(\sigma_3) )
\nonumber \\
&& + ( C_{ij}(\sigma_1) C_{jk}(\sigma_2) - V_{ij}(\sigma_1) \cdot V_{jk}(\sigma_2) ) ( C_{jk}(\sigma_2) C_{ki}(\sigma_3) - V_{jk}(\sigma_2) \cdot V_{ki}(\sigma_3) ) 
\},
\end{eqnarray}
\begin{eqnarray}
P_4 &=& 
-4 ( V_{ij}(\sigma_1) \cdot r_{ij}(\tau_1) ) ( V_{ij}(\sigma_1) \cdot r_{ij}(\tau_2) )
\{ 2 + ( C_{jk}(\sigma_2) C_{ki}(\sigma_3) - V_{jk}(\sigma_2) \cdot V_{ki}(\sigma_3) )^2 \} 
\nonumber \\
&& +8 ( V_{jk}(\sigma_2) \cdot r_{ij}(\tau_1) ) ( V_{jk}(\sigma_2) \cdot r_{ij}(\tau_2) )
\{ 1 + ( C_{ij}(\sigma_1) C_{ki}(\sigma_3) - V_{ij}(\sigma_1) \cdot V_{ki}(\sigma_3) )^2 \} 
\nonumber \\
&& +8 ( V_{ki}(\sigma_3) \cdot r_{ij}(\tau_1) ) ( V_{ki}(\sigma_3) \cdot r_{ij}(\tau_2) )
\{ 1 + ( C_{ij}(\sigma_1) C_{jk}(\sigma_2)- V_{ij}(\sigma_1) \cdot V_{jk}(\sigma_2) )^2 \} 
\nonumber \\
&& +4 ( V_{ij}(\sigma_1) \cdot V_{ki}(\sigma_3) )
\{ V_{ij}(\sigma_1) \cdot ( r_{ij}(\tau_1) -r_{ki}(\tau_1) ) \} \{ V_{ki}(\sigma_3) \cdot ( r_{jk}(\tau_2) - r_{ij}(\tau_2) ) \} 
\nonumber \\
&& +4 ( V_{ij}(\sigma_1) \cdot V_{jk}(\sigma_2) -2 C_{ij}(\sigma_1) C_{jk}(\sigma_2) )
\nonumber \\ && 
\qquad \times
\{ V_{jk}(\sigma_2) \cdot ( r_{ij}(\tau_1) -r_{ki}(\tau_1) ) \} \{ V_{ij}(\sigma_1) \cdot ( r_{jk}(\tau_2) - r_{ij}(\tau_2) ) \} 
\nonumber \\
&& +4 \{ -C_{jk}(\sigma_2) C_{ki}(\sigma_3) + V_{jk}(\sigma_2) \cdot V_{ki}(\sigma_3) + 2 C_{ij}(\sigma_1)^2 C_{jk}(\sigma_2) C_{ki}(\sigma_3)
\nonumber \\
&& +2 ( V_{ij}(\sigma_1) \cdot V_{jk}(\sigma_2) ) ( V_{ij}(\sigma_1) \cdot V_{ki}(\sigma_3) )
-4 C_{ij}(\sigma_1) C_{jk}(\sigma_2) ( V_{ij}(\sigma_1) \cdot V_{ki}(\sigma_3) ) \} 
\nonumber \\
&& \qquad \times
\{ V_{jk}(\sigma_2) \cdot ( r_{ij}(\tau_1) -r_{ki}(\tau_1) ) \} \{ V_{ki}(\sigma_3) \cdot ( r_{jk}(\tau_2) - r_{ij}(\tau_2) ) \} 
\nonumber \\
&& +32 \{
-2 ( V_{jk}(\sigma_2) \cdot r_{jk}(\tau_1) ) ( V_{ki}(\sigma_3) \cdot r_{ki}(\tau_2) )
- ( V_{ki}(\sigma_3) \cdot r_{jk}(\tau_1) ) ( V_{jk}(\sigma_2) \cdot r_{ki}(\tau_2) ) 
\nonumber \\
&& + ( C_{jk}(\sigma_2) C_{ki}(\sigma_3) - V_{jk}(\sigma_2) \cdot V_{ki}(\sigma_3) )
( V_{ij}(\sigma_1) \cdot r_{jk}(\tau_1) ) ( V_{ij}(\sigma_1) \cdot r_{ki}(\tau_2) ) 
\nonumber \\
&& - ( C_{ij}(\sigma_1) C_{jk}(\sigma_2) - V_{ij}(\sigma_1) \cdot V_{jk}(\sigma_2) )
( V_{ki}(\sigma_3) \cdot r_{jk}(\tau_1) ) ( V_{ij}(\sigma_1) \cdot r_{ki}(\tau_2) ) 
\nonumber \\
&& - ( C_{ij}(\sigma_1) C_{ki}(\sigma_3) - V_{ij}(\sigma_1) \cdot V_{ki}(\sigma_3) )
( V_{ij}(\sigma_1) \cdot r_{jk}(\tau_1) ) ( V_{jk}(\sigma_2) \cdot r_{ki}(\tau_2) )
\}.
\end{eqnarray}
\section{Examples of evaluations of P's}
\setcounter{equation}{0}
In this appendix, we present some examples
of evaluations of $\tilde{\Gamma}_{(2)}$ from 
individual terms in (\ref{Gamma2}).
For each example below, we first write down
the whole expression of the one-particle irreducible, planar contractions, and 
then present the final form which fits into the 
form (\ref{Pform}).

The first two terms in (\ref{Gamma2}) involve the quartic vertices
and contribute to $P_q$.
\begin{eqnarray}
&& \frac14 g^2 \kappa
\int d\tau \langle {\rm tr} [Y^n(\tau),Y^m(\tau)] [Y^n(\tau),Y^m(\tau)] \rangle_{1PI,{\rm planar}} \nonumber \\ &&
= \frac12 g^2 \kappa  \int d\tau \sum_{i,j,k} (
	\langle Y^n_{ij}(\tau) Y^m_{ji}(\tau) \rangle_0~
	\langle Y^n_{ik}(\tau) Y^m_{ki}(\tau) \rangle_0~
-	\langle Y^n_{ij}(\tau) Y^n_{ji}(\tau) \rangle_0~
	\langle Y^m_{ik}(\tau) Y^m_{ki}(\tau) \rangle_0~ ) 
\nonumber \\ &&
= \frac{g^2}{\kappa} \int d \tau_1 d \tau_2 d\sigma_1 d\sigma_2 d\sigma_3
\sum_{i,j,k} 
\{ -36 -8 V_{jk}(\sigma_2)^2 -8 V_{ki}(\sigma_3)^2
-2 V_{jk}(\sigma_2)^2 V_{ki}(\sigma_3)^2 
\nonumber \\ && \qquad \quad 
+ 2 ( V_{jk}(\sigma_2) \cdot V_{ki}(\sigma_3) )^2 \} 
\{ -\partial_{\sigma_1} \Delta_{ij} (\sigma_1,\tau_1,\tau_2) \}
\Delta_{jk} (\sigma_2,\tau_1,\tau_2) 
\Delta_{ki} (\sigma_3,\tau_1,\tau_2),
\end{eqnarray}
\begin{eqnarray}
&& \frac12 g^2 \kappa
\int d\tau \langle {\rm tr} [\tilde{A}(\tau),Y^m(\tau)] [\tilde{A}(\tau),Y^m(\tau)] \rangle_{1PI,{\rm planar}} \nonumber \\ &&
= g^2 \kappa  \int d\tau \sum_{i,j,k} (
	\langle \tilde{A}_{ij}(\tau) Y^m_{ji}(\tau) \rangle_0~
	\langle \tilde{A}_{ik}(\tau) Y^m_{ki}(\tau) \rangle_0~
-	\langle \tilde{A}_{ij}(\tau) \tilde{A}_{ji}(\tau) \rangle_0~
	\langle Y^m_{ik}(\tau) Y^m_{ki}(\tau) \rangle_0~ 
\nonumber \\ && \qquad \qquad \qquad 
+	\langle \tilde{A}_{ij}(\tau) Y^m_{ji}(\tau) \rangle_0~
	\langle Y^m_{jk}(\tau) \tilde{A}_{kj}(\tau) \rangle_0~
-	\langle \tilde{A}_{ij}(\tau) Y^m_{ji}(\tau) \rangle_0~
	\langle \tilde{A}_{jk}(\tau) Y^m_{kj}(\tau) \rangle_0~ )
\nonumber \\ &&
= \frac{g^2}{\kappa} \int d \tau_1 d \tau_2 d\sigma_1 d\sigma_2 d\sigma_3
\sum_{i,j,k}
\{ -9 -10 V_{jk}(\sigma_2)^2 -10 V_{ki}(\sigma_3)^2
\nonumber \\ && \qquad \qquad 
+ 12 C_{jk}(\sigma_2) C_{ki}(\sigma_3) ( V_{jk}(\sigma_2) \cdot V_{ki}(\sigma_3) )
-4 V_{jk}(\sigma_2)^2 V_{ki}(\sigma_3)^2 \} 
\nonumber \\ && \qquad \qquad \qquad 
\times
\{ -\partial_{\sigma_1} \Delta_{ij} (\sigma_1,\tau_1,\tau_2) \}
\Delta_{jk} (\sigma_2,\tau_1,\tau_2) 
\Delta_{ki} (\sigma_3,\tau_1,\tau_2). 
\end{eqnarray}
There are three terms which contain ghost fields.
One of them involves two derivatives, thus contributes 
to the $P_1$ term:
\begin{eqnarray}
&& -\frac12 g^2 \kappa^2
\int d\tau_1 d\tau_2 
\langle {\rm tr} \partial_{\tau_1} \bar{c}(\tau_1)
[\tilde{A}(\tau_1),c(\tau_1)] 
~{\rm tr} \partial_{\tau_2} \bar{c}(\tau_2)[\tilde{A}(\tau_2),c(\tau_2)] \rangle_{1PI,{\rm planar}} \nonumber \\ &&
= -\frac12 g^2 \kappa^2 \int d\tau_1 d\tau_2 \sum_{i,j,k} (
           \langle \tilde{A}_{ij}(\tau_1) \tilde{A}_{ji}(\tau_2) \rangle_0~
           \langle c_{jk}(\tau_1) \partial_{\tau_2} \bar{c}_{kj}(\tau_2) \rangle_0~
           \langle \partial_{\tau_1} \bar{c}_{ki}(\tau_1) c_{ik}(\tau_2) \rangle_0~ 
\nonumber \\ && \qquad \qquad \qquad \qquad \qquad 
+          \langle \tilde{A}_{ij}(\tau_1) \tilde{A}_{ji}(\tau_2) \rangle_0~
           \langle \partial_{\tau_1} \bar{c}_{jk}(\tau_1) c_{kj}(\tau_2) \rangle_0~
           \langle c_{ki}(\tau_1) \partial_{\tau_2} \bar{c}_{ik}(\tau_2) \rangle_0~
            ) \nonumber \\ &&
= \frac{g^2}{\kappa} \int d \tau_1 d \tau_2 d\sigma_1 d\sigma_2 d\sigma_3
\sum_{i,j,k}
\left( -\frac{1}{2} + V_{ij}(\sigma_1)^2 - V_{jk}(\sigma_2)^2 - V_{ki}(\sigma_3)^2 \right) \nonumber \\ && \qquad \qquad \qquad 
\times \{ \partial_{\tau_1} \partial_{\tau_2}
\Delta_{ij} (\sigma_1,\tau_1,\tau_2) \}
\Delta_{jk} (\sigma_2,\tau_1,\tau_2) 
\Delta_{ki} (\sigma_3,\tau_1,\tau_2). 
\end{eqnarray}
The term below has one derivative and contribute to $P_2$.
\begin{eqnarray}
&& ig^2 \kappa^2
\int d\tau_1 d\tau_2 
\langle {\rm tr} 
\partial_{\tau_1} \bar{c}(\tau_1)[\tilde{A}(\tau_1),c(\tau_1)] 
~{\rm tr} [B^m,\bar{c}](\tau_2)[Y^m(\tau_2),c(\tau_2)] 
\rangle_{1PI,{\rm planar}} 
\nonumber \\ && 
= ig^2 \kappa^2 \int d\tau_1 d\tau_2 \sum_{i,j,k} (
           \langle \tilde{A}_{ij}(\tau_1) Y^m_{ji}(\tau_2) \rangle_0~
           \langle c_{jk}(\tau_1) [B^m,\bar{c}]_{kj}(\tau_2) \rangle_0~
           \langle \partial_{\tau_1} \bar{c}_{ki}(\tau_1) c_{ik}(\tau_2) \rangle_0~ \nonumber \\ &&  \qquad \qquad \qquad \qquad
+          \langle \tilde{A}_{ij}(\tau_1) Y^m_{ji}(\tau_2) \rangle_0~
           \langle \partial_{\tau_1} \bar{c}_{jk}(\tau_1) c_{kj}(\tau_2) \rangle_0~
           \langle c_{ki}(\tau_1) [B^m,\bar{c}]_{ik}(\tau_2) \rangle_0~ ) 
\nonumber \\ && 
= \frac{g^2}{\kappa} \int d \tau_1 d \tau_2 d\sigma_1 d\sigma_2 d\sigma_3
\sum_{i,j,k}
\{ 2 C_{ij}(\sigma_1) ( V_{ij}(\sigma_1) \cdot r_{ij}(\tau_2) )
\nonumber \\ && \qquad \qquad
-2 C_{jk}(\sigma_2) ( V_{jk}(\sigma_2) \cdot r_{ij}(\tau_2) )
-2 C_{ki}(\sigma_3) ( V_{ki}(\sigma_3) \cdot r_{ij}(\tau_2) ) \} 
\nonumber \\ &&  \qquad \qquad \qquad
\times
\{ \partial_{\tau_1}
\Delta_{ij} (\sigma_1,\tau_1,\tau_2) \}
\Delta_{jk} (\sigma_2,\tau_1,\tau_2) 
\Delta_{ki} (\sigma_3,\tau_1,\tau_2). 
\end{eqnarray}
The last one involves no derivatives and
 contributes to both $P_3$ and $P_4$.
\begin{eqnarray}
&& \frac12 g^2 \kappa^2
\int d\tau_1 d\tau_2 
\langle {\rm tr} [B^n,\bar{c}](\tau_1)[Y^n(\tau_1),c(\tau_1)] 
~{\rm tr} [B^m,\bar{c}](\tau_2)[Y^m(\tau_2),c(\tau_2)] \rangle_{1PI,{\rm planar}} \nonumber \\ &&
= \frac12 g^2 \kappa^2 \int d\tau_1 d\tau_2 \sum_{i,j,k} (
           \langle Y^n_{ij}(\tau_1) Y^m_{ji}(\tau_2) \rangle_0~
           \langle c_{jk}(\tau_1) [B^m,\bar{c}]_{kj}(\tau_2) \rangle_0~
           \langle [B^n,\bar{c}]_{ki}(\tau_1) c_{ik}(\tau_2) \rangle_0~ 
\nonumber \\ && \qquad \qquad \qquad \qquad 
+          \langle Y^n_{ij}(\tau_1) Y^m_{ji}(\tau_2) \rangle_0~
           \langle [B^n,\bar{c}]_{jk}(\tau_1) c_{kj}(\tau_2) \rangle_0~
           \langle c_{ki}(\tau_1) [B^m,\bar{c}]_{ik}(\tau_2) \rangle_0~ ) \nonumber \\ &&
= \frac{g^2}{\kappa} \int d \tau_1 d \tau_2 d\sigma_1 d\sigma_2 d\sigma_3
\sum_{i,j,k}
\left\{ \right.
-\frac{1}{2} ( r_{ij}(\tau_1) \cdot r_{ij}(\tau_2) )
+ ( V_{ij}(\sigma_1) \cdot r_{ij}(\tau_1) ) ( V_{ij}(\sigma_1) \cdot r_{ij}(\tau_2) ) 
\nonumber \\ && \qquad \qquad 
- ( V_{jk}(\sigma_2) \cdot r_{ij}(\tau_1) ) ( V_{jk}(\sigma_2) \cdot r_{ij}(\tau_2) )
- ( V_{ki}(\sigma_3) \cdot r_{ij}(\tau_1) ) ( V_{ki}(\sigma_3) \cdot r_{ij}(\tau_2) )
\left. \right\} 
\nonumber \\ && \qquad \qquad \qquad 
\times \Delta_{ij} (\sigma_1,\tau_1,\tau_2) 
\Delta_{jk} (\sigma_2,\tau_1,\tau_2) 
\Delta_{ki} (\sigma_3,\tau_1,\tau_2). 
\end{eqnarray}

There are six terms which consist only of
$\tilde{A}$ and $Y^n$.
Each of these terms has twelve ways of one-particle irreducible,
planar contractions and 
is much more complicated than the ghost terms.
However,  they can be arranged 
into the form (\ref{Pform}) by similar transformations.

The final example is the terms with fermion fields.
The three terms with fermion fields give 
contributions to $P_1$, $P_2$, $P_3$, and $P_4$.
\begin{eqnarray}
&& -\frac18 g^2 \kappa^2 
\int d\tau_1 d\tau_2 
\langle {\rm tr} \theta^\alpha(\tau_1)
[\tilde{A}(\tau_1),\theta^\alpha(\tau_1)] 
~{\rm tr} \theta^\beta(\tau_2)[\tilde{A}(\tau_2),\theta^\beta(\tau_2)] \rangle_{1PI,{\rm planar}} \nonumber \\ &&
= -\frac12 g^2 \kappa^2 \int d\tau_1 d\tau_2 \sum_{i,j,k} 
           \langle \tilde{A}_{ij}(\tau_1) \tilde{A}_{ji}(\tau_2) \rangle_0~
           \langle \theta^\alpha_{jk}(\tau_1) \theta^\beta_{kj}(\tau_2) \rangle_0~
           \langle \theta^\alpha_{ki}(\tau_1) \theta^\beta_{ik}(\tau_2) \rangle_0~ \nonumber \\ &&
= \frac{g^2}{\kappa} \int d \tau_1 d \tau_2 d\sigma_1 d\sigma_2 d\sigma_3
\sum_{i,j,k} \left[ \right.
4 \{
C_{jk}(\sigma_2) C_{ki}(\sigma_3) - V_{jk}(\sigma_2) \cdot V_{ki}(\sigma_3) 
\nonumber \\ && \qquad \qquad
+2 V_{ij}(\sigma_1)^2 C_{jk}(\sigma_2) C_{ki}(\sigma_3) -2 V_{ij}(\sigma_1)^2 ( V_{jk}(\sigma_2) \cdot V_{ki}(\sigma_3) ) 
\nonumber \\ && \qquad \qquad
- C_{ij}(\sigma_1) C_{ki}(\sigma_3) + V_{ij}(\sigma_1) \cdot V_{ki}(\sigma_3) 
\nonumber \\ && \qquad \qquad
-2 V_{jk}(\sigma_2)^2 C_{ij}(\sigma_1) C_{ki}(\sigma_3) +2 V_{jk}(\sigma_2)^2 ( V_{ij}(\sigma_1) \cdot V_{ki}(\sigma_3) ) 
\nonumber \\ && \qquad \qquad
- C_{ij}(\sigma_1) C_{jk}(\sigma_2) + V_{ij}(\sigma_1) \cdot V_{jk}(\sigma_2) 
\nonumber \\ && \qquad \qquad
-2 V_{ki}(\sigma_3)^2 C_{ij}(\sigma_1) C_{jk}(\sigma_2) +2 V_{ki}(\sigma_3)^2 ( V_{ij}(\sigma_1) \cdot V_{jk}(\sigma_2) ) \} 
\nonumber \\ && \qquad \qquad \qquad \qquad
\times \{ \partial_{\tau_1} \partial_{\tau_2}
\Delta_{ij} (\sigma_1,\tau_1,\tau_2) \}
\Delta_{jk} (\sigma_2,\tau_1,\tau_2) 
\Delta_{ki} (\sigma_3,\tau_1,\tau_2) 
\nonumber \\ && \qquad \qquad
+ 16 \{ C_{ij}(\sigma_1) ( V_{jk}(\sigma_2) \cdot r_{jk}(\tau_2) ) - C_{jk}(\sigma_2) ( V_{ij}(\sigma_1) \cdot r_{jk}(\tau_2) )
\nonumber \\ && \qquad \qquad
+2 C_{ij}(\sigma_1) V_{ki}(\sigma_3)^2 ( V_{jk}(\sigma_2) \cdot r_{jk}(\tau_2) ) 
-2 C_{jk}(\sigma_2) V_{ki}(\sigma_3)^2 ( V_{ij}(\sigma_1) \cdot r_{jk}(\tau_2) ) \}
\nonumber \\ && \qquad \qquad \qquad \qquad
\times 
\{ \partial_{\tau_1}
\Delta_{ij} (\sigma_1,\tau_1,\tau_2) \}
\Delta_{jk} (\sigma_2,\tau_1,\tau_2) 
\Delta_{ki} (\sigma_3,\tau_1,\tau_2) 
\nonumber \\ && \qquad \qquad
+ 4 \{
- C_{jk}(\sigma_2) C_{ki}(\sigma_3) - V_{jk}(\sigma_2) \cdot V_{ki}(\sigma_3) 
\nonumber \\ && \qquad \qquad
-2 V_{ij}(\sigma_1)^2 C_{jk}(\sigma_2) C_{ki}(\sigma_3) -2 V_{ij}(\sigma_1)^2 ( V_{jk}(\sigma_2) \cdot V_{ki}(\sigma_3) ) 
\nonumber \\ && \qquad \qquad
+ C_{ij}(\sigma_1) C_{ki}(\sigma_3) + V_{ij}(\sigma_1) \cdot V_{ki}(\sigma_3) 
\nonumber \\ && \qquad \qquad
+2 V_{jk}(\sigma_2)^2 C_{ij}(\sigma_1) C_{ki}(\sigma_3) +2 V_{jk}(\sigma_2)^2 ( V_{ij}(\sigma_1) \cdot V_{ki}(\sigma_3) ) 
\nonumber \\ && \qquad \qquad
+ C_{ij}(\sigma_1) C_{jk}(\sigma_2) + V_{ij}(\sigma_1) \cdot V_{jk}(\sigma_2)
\nonumber \\ && \qquad \qquad
+2 V_{ki}(\sigma_3)^2 C_{ij}(\sigma_1) C_{jk}(\sigma_2) +2 V_{ki}(\sigma_3)^2 ( V_{ij}(\sigma_1) \cdot V_{jk}(\sigma_2) ) \} 
\nonumber \\ && \qquad \qquad \qquad \qquad
\times
( r_{ij}(\tau_1) \cdot r_{ij}(\tau_2) )
\Delta_{ij} (\sigma_1,\tau_1,\tau_2) 
\Delta_{jk} (\sigma_2,\tau_1,\tau_2) 
\Delta_{ki} (\sigma_3,\tau_1,\tau_2) 
\nonumber \\ && \qquad \qquad
+ 8 \{
( V_{jk}(\sigma_2) \cdot r_{jk}(\tau_1) ) ( V_{ki}(\sigma_3) \cdot r_{ki}(\tau_2) )
+ ( V_{ki}(\sigma_3) \cdot r_{jk}(\tau_1) ) ( V_{jk}(\sigma_2) \cdot r_{ki}(\tau_2) ) 
\nonumber \\ && \qquad \qquad
+2 V_{ij}(\sigma_1)^2 ( V_{jk}(\sigma_2) \cdot r_{jk}(\tau_1) ) ( V_{ki}(\sigma_3) \cdot r_{ki}(\tau_2) )
\nonumber \\ && \qquad \qquad
+2 V_{ij}(\sigma_1)^2 ( V_{ki}(\sigma_3) \cdot r_{jk}(\tau_1) ) ( V_{jk}(\sigma_2) \cdot r_{ki}(\tau_2) )
\} \nonumber \\ && \qquad \qquad \qquad \qquad
\times
\Delta_{ij} (\sigma_1,\tau_1,\tau_2) 
\Delta_{jk} (\sigma_2,\tau_1,\tau_2) 
\Delta_{ki} (\sigma_3,\tau_1,\tau_2) 
\left. \right].
\end{eqnarray}
\section{Formulas in calculating $\tilde{\Gamma}_{Y}$}
\setcounter{equation}{0}
Here we supplement the derivation of $\tilde{\Gamma}_{Y}$
by providing some formulas. From 
the explicit form of the proper-time propagator (\ref{Delta}),
we have
\begin{eqnarray}
&& \partial_{\tau_1}
\Delta_{ij} (\sigma_1,\tau_1,\tau_2)
\nonumber \\
&& \qquad = \left[
- \frac{V_{ij}(\sigma_1)}{C_{ij}(\sigma_1)} \cdot \frac{r_{ij}(\tau_1)+r_{ij}(\tau_2)}{2}
- \frac{C_{ij}(\sigma_1)}{V_{ij}(\sigma_1)} \frac{|r_{ij}(\tau_1)-r_{ij}(\tau_2)|}{2}
\right] 
\Delta_{ij} (\sigma_1,\tau_1,\tau_2), 
\nonumber \\
&& \partial_{\tau_1} \partial_{\tau_2}
\Delta_{ij} (\sigma_1,\tau_1,\tau_2) 
\nonumber \\
&& \qquad = \left[
-\frac{\tilde{v}_{ij}}{2} \frac{V_{ij}(\sigma_1)}{C_{ij}(\sigma_1)} 
+\frac{\tilde{v}_{ij}}{2} \frac{C_{ij}(\sigma_1)}{V_{ij}(\sigma_1)}
+ \left( \frac{V_{ij}(\sigma_1)}{C_{ij}(\sigma_1)} \cdot \frac{r_{ij}(\tau_1)+r_{ij}(\tau_2)}{2} \right)^2 \right.
\nonumber \\
&& \qquad \qquad \left.
- \left( \frac{C_{ij}(\sigma_1)}{V_{ij}(\sigma_1)} \frac{r_{ij}(\tau_1)-r_{ij}(\tau_2)}{2} \right)^2
\right] 
\Delta_{ij} (\sigma_1,\tau_1,\tau_2), 
\nonumber \\
&& \frac{1}{2} ( \partial_{\tau_1}^2 + \partial_{\tau_2}^2 )
\Delta_{ij} (\sigma_1,\tau_1,\tau_2)
\nonumber \\
&& \qquad = \left[
-\frac{\tilde{v}_{ij}}{2} \frac{V_{ij}(\sigma_1)}{C_{ij}(\sigma_1)} 
-\frac{\tilde{v}_{ij}}{2} \frac{C_{ij}(\sigma_1)}{V_{ij}(\sigma_1)}
+ \left( \frac{V_{ij}(\sigma_1)}{C_{ij}(\sigma_1)} \cdot \frac{r_{ij}(\tau_1)+r_{ij}(\tau_2)}{2} \right)^2 \right.
\nonumber \\
&& \qquad \qquad \left.
+ \left( \frac{C_{ij}(\sigma_1)}{V_{ij}(\sigma_1)} \frac{r_{ij}(\tau_1)-r_{ij}(\tau_2)}{2} \right)^2
\right]
\Delta_{ij} (\sigma_1,\tau_1,\tau_2).
\end{eqnarray}
In general, $P_Y$ (\ref{Py}) has complicated dependence on $\tau_1$ and
$\tau_2$. 
The dependence on $\{ x \cdot \tilde{v} \}$
in proper-time propagators further complicates them.
When we restrict to the case where
$\{ x \cdot \tilde{v} \}$ vanish, however, 
all we need is to notice the following equalities:
\begin{eqnarray}
&& \int d \tau_1 d \tau_2
\left( \frac{\tau_1+\tau_2}{2} \right)^2
\Delta_{ij} (\sigma_1,\tau_1,\tau_2) 
\Delta_{jk} (\sigma_2,\tau_1,\tau_2) 
\Delta_{ki} (\sigma_3,\tau_1,\tau_2) 
\nonumber \\
&& \qquad = \int d \tau_1 d \tau_2~
\frac{1}{2} \left( 
\frac{\tilde{v}_{ij} V_{ij}(\sigma_1)}{C_{ij}(\sigma_1)} + \frac{\tilde{v}_{jk} V_{jk}(\sigma_2)}{C_{jk}(\sigma_2)} + \frac{\tilde{v}_{ki} V_{ki}(\sigma_3)}{C_{ki}(\sigma_3)}
\right)^{-1}
\nonumber \\
&& \qquad \qquad \times \Delta_{ij} (\sigma_1,\tau_1,\tau_2) 
\Delta_{jk} (\sigma_2,\tau_1,\tau_2) 
\Delta_{ki} (\sigma_3,\tau_1,\tau_2), 
\nonumber \\
&& \int d \tau_1 d \tau_2
\left( \frac{\tau_1-\tau_2}{2} \right)^2
\Delta_{ij} (\sigma_1,\tau_1,\tau_2) 
\Delta_{jk} (\sigma_2,\tau_1,\tau_2) 
\Delta_{ki} (\sigma_3,\tau_1,\tau_2) 
\nonumber \\
&& \qquad = \int d \tau_1 d \tau_2~
\frac{1}{2} \left( 
\frac{\tilde{v}_{ij} C_{ij}(\sigma_1)}{V_{ij}(\sigma_1)} + \frac{\tilde{v}_{jk} C_{jk}(\sigma_2)}{V_{jk}(\sigma_2)} + \frac{\tilde{v}_{ki} C_{ki}(\sigma_3)}{V_{ki}(\sigma_3)}
\right)^{-1}
\nonumber \\
&& \qquad \qquad \times \Delta_{ij} (\sigma_1,\tau_1,\tau_2) 
\Delta_{jk} (\sigma_2,\tau_1,\tau_2) 
\Delta_{ki} (\sigma_3,\tau_1,\tau_2). 
\end{eqnarray}


 \end{document}